\documentclass[preprint,12pt, 
aps,amsmath,amssymb,nofootinbib,superscriptaddress,hyperref,
]{revtex4} 
\usepackage[utf8]{inputenc}
\usepackage{epsfig} 
\usepackage{wasysym} 
\usepackage{mathrsfs} 
\usepackage{graphicx} 
\usepackage{amsfonts} 
\usepackage{amsbsy} 
\usepackage{amscd} 
\usepackage{pstricks} 
\usepackage{multirow}
\usepackage{slashed} 
\usepackage{tikz}
\usetikzlibrary{decorations.pathmorphing,decorations.markings}
\newcommand{\beq}{\begin{equation}}
\newcommand{\eeq}{\end{equation}}
\newcommand{\bea}{\begin{eqnarray}}
\newcommand{\eea}{\end{eqnarray}}
\newcommand{\beas}{\begin{eqnarray*}}
\newcommand{\eeas}{\end{eqnarray*}}
\newcommand{\bi}{\begin{itemize}}
\newcommand{\ei}{\end{itemize}}

\def\tev{\,{\ifmmode\mathrm {TeV}\else TeV\fi}}
\def\gev{\,{\ifmmode\mathrm {GeV}\else GeV\fi}}
\def\to{\rightarrow}

\allowdisplaybreaks

\begin{document}

\title{
Bounds on heavy Majorana neutrinos in type-I seesaw \\ 
and implications for collider searches 
}

\author{Arindam Das\footnote{arindam@kias.re.kr}}
\affiliation{School of Physics, KIAS, Seoul 130-722, Korea}
\affiliation{Department of Physics \& Astronomy, Seoul National University 1 Gwanak-ro, Gwanak-gu, Seoul 08826, Korea}
\affiliation{Korea Neutrino Research Center, Bldg 23-312, Seoul National University, Sillim-dong, Gwanak-gu, Seoul 08826, Korea}
\author{Nobuchika Okada\footnote{okadan@ua.edu}}
\affiliation{Department of Physics and Astronomy, 
University of Alabama, \\ 
Tuscaloosa,  Alabama 35487, USA 
}

\begin{abstract}

The neutrino masses and flavor mixings, which are missing in the Standard Model (SM), can be 
  naturally incorporated in the type-I seesaw extension of the SM 
  with heavy Majorana neutrinos being singlet under the SM gauge group. 
If the heavy Majorana neutrinos are around the electroweak scale and their mixings with the SM neutrinos 
  are sizable, they can be produced at high energy colliders, leaving characteristic signatures 
  with lepton-number violations.  
Employing the general parametrization for the neutrino Dirac mass matrix in the minimal seesaw scenario, 
  we perform a parameter scan and identify allowed regions to satisfy a variety of experimental constraints 
  from the neutrino oscillation data, the electroweak precision measurements and the lepton-flavor violating processes. 
We find that the resultant mixing parameters between the heavy neutrinos and the SM neutrinos 
  are more severely constrained than those obtained from the current search for heavy Majorana neutrinos at the LHC. 
Such parameter regions can be explored at the High-Luminosity LHC and a 100 TeV pp-collider in the future. 

\end{abstract}


\maketitle

With the measurements of nonzero reactor angle $\theta_{13}$ \cite{Neut1, Neut2, Neut4, Neut5, Neut6}, 
   all neutrino oscillation data expect the Dirac $CP$-phase have been determined \cite{Neut3},  
   which indicate physics beyond the Standard Model (SM). 
The type-I seesaw extension \cite{seesaw0, seesaw1, seesaw2, seesaw3, seesaw4, seesaw5, seesaw6}  
   of the SM is arguably the simplest idea to naturally incorporate the tiny neutrino masses 
   and the flavor mixings into the SM, 
   where heavy Majorana neutrinos which are singlet under the SM gauge group are introduced. 
The heavy neutrinos are integrated out at low energies, leading to a dimension five operator \cite{Weinberg:1979sa} 
   among the SM lepton and the Higgs doublets at low energies. 
After the electroweak symmetry breaking, light Majorana masses for the SM neutrinos are generated  
   thought the type-I seesaw mechanism.

Although the heavy Majorana neutrinos are singlet under the SM gauge group, 
   the heavy mass eigenstates after the seesaw mechanism couple with the weak bosons 
   and the Higgs boson through the mixing with the SM neutrinos. 
If the heavy neutrinos are around or below the electroweak scale and the mixing with the SM neutrinos 
   is not extremely small, the heavy Majorana neutrinos can be produced at high energy colliders.  
The smoking gun collider signature of heavy neutrino production at the collider experiments 
   is the same-sign dilepton in the final state which reflects the lepton-number violation 
   due to their Majorana masses.  
The heavy neutrino signature, once observed at collider experiments, 
   can provide us with a clue to explore the origin of the neutrino masses and flavor mixings.

The mixing of the heavy neutrinos with the SM neutrinos affects not only the production cross section 
  at high energy colliders but also a variety of phenomenologies such as 
  the neutrino oscillation data \cite{Escrihuela:2015wra, Ge:2016xya}, 
  the precision measurement of weak boson decays, 
  and the lepton-flavor-violating decays of charged leptons 
\cite{Constraints1, Constraints2, Constraints3, Constraints4, Constraints5} 
\cite{Asaka:2011pb, Ruchayskiy:2011aa, Gorbunov:2014ypa, Drewes:2015iva,Hernandez:2016kel,Drewes:2016jae} 
\cite{Abada:2013aba, Abada:2014kba, Asaka:2014kia, Fernandez-Martinez:2015hxa, deGouvea:2015euy, Fernandez-Martinez:2016lgt}, 
  which severely constrain the mixing parameters. 
Therefore, in order to discuss the possibility of the heavy neutrino production at high energy colliders, 
  it is essential to identify allowed regions for the mixing parameters 
  from the current phenomenological constraints. 
In this letter, for simplicity, we consider the minimal seesaw scenario \cite{minimal_seesaw1, minimal_seesaw2} 
   and introduce two right-handed neutrinos to the SM, 
   which is the minimal setup to reproduce the observed neutrino oscillation data 
   with a prediction of one massless neutrino. 
Employing the general parametrization for the neutrino Dirac mass matrix in the seesaw model, 
  we perform a parameter scan to identify the allowed regions for the mixing parameters.

Let us begin with a brief review of the minimal seesaw. 
We introduce two flavors of right-handed neutrinos $N_R^{j}$ ($j=1,2$).  
The relevant part of the Lagrangian is written as
\bea
\mathcal{L} \supset -\sum_{i=1}^3 \sum_{j=1}^2 Y_D^{ij} \overline{\ell_L^{~i}}H N_R^{j} 
                   -\frac{1}{2} \sum_{k=1}^2 m_N^{~k} \overline{N_R^{k C}} N_R^k  + \rm{H. c.} , 
\label{typeI}
\eea
where $\ell_{L}^{~i}$ ($i=1, 2, 3$) and $H$ are the SM lepton doublet of the $i$-th generation 
  and the SM Higgs doublet, respectively, and the Majorana mass matrix of the right-handed neutrinos
  is taken to be diagonal without loss of generality.  
After the electroweak symmetry breaking, we obtain the Dirac mass matrix as 
  $m_{D}= \frac{Y_D}{\sqrt{2}} v$, where $v=246$ GeV is the Higgs vacuum expectation value. 
Using the Dirac and Majorana mass matrices, the neutrino mass matrix is expressed as  
\bea
{\cal M}_{\nu}=\begin{pmatrix}
0&&m_{D}\\
m_{D}^{T}&&m_{N}
\end{pmatrix}.
\label{typeInu}
\eea
Assuming the hierarchy of $|m_D^{ij}/m_N^{~k}| \ll 1$, we diagonalize the mass matrix and obtain 
  the seesaw formula for the light Majorana neutrinos as 
\bea
m_{\nu} \simeq - m_{D} m_{N}^{-1} m_{D}^{T}.
\label{seesawI}
\eea
We express the light neutrino flavor eigenstate $(\nu)$ in terms of the mass eigenstates 
  of the light $(\nu_m)$ and heavy $(N_m)$ Majorana neutrinos such as 
$\nu \simeq \mathcal{N} \nu_m+\mathcal{R} N_m$, 
  where $\mathcal{R} =m_D m_N^{-1}$, $\mathcal{N}=\Big(1-\frac{1}{2}\epsilon\Big) U_{\rm{MNS}}$ 
    with $\epsilon=\mathcal{R}^\ast\mathcal{R}^T$ and $U_{\rm{MNS}}$ is the neutrino mixing matrix 
   which diagonalizes the light neutrino mass matrix as
\bea
U_{\rm{MNS}}^T m_\nu U_{\rm{MNS}} ={\rm diag}(m_1, m_2, m_3).
\eea
In the presence of $\epsilon$, the mixing matrix $\mathcal{N}$ is not unitary, namely $\mathcal{N}^\dagger\mathcal{N}\neq1$.

In terms of the neutrino mass eigenstates, the charged current interaction can be written as 
\bea 
\mathcal{L}_{CC}= 
 -\frac{g}{\sqrt{2}} W_{\mu}
  \overline{\ell_\alpha} \gamma^{\mu} P_L 
   \left( {\cal N}_{\alpha j} \nu_{m_j}+ {\cal R}_{\alpha j} N_{m_j} \right) + \rm{H.c.}, 
\label{CC}
\eea
where $\ell_\alpha$ ($\alpha=e, \mu, \tau$) denotes the three generations of the charged leptons, 
  and $P_L =  (1- \gamma_5)/2$. 
Similarly, the neutral current interaction is given by 
\bea 
\mathcal{L}_{NC}&=& 
 -\frac{g}{2 \cos \theta_{\rm W}}  Z_{\mu} 
\Big[ 
  \overline{\nu_{m_i}} \gamma^{\mu} P_L ({\cal N}^\dagger {\cal N})_{ij} \nu_{m_{j}}
 +  \overline{N_{m_{i}}} \gamma^{\mu} P_L ({\cal R}^\dagger {\cal R})_{ij} N_{m_{j}} \nonumber \\
&+& \Big\{ 
  \overline{\nu_{m_{i}}} \gamma^{\mu} P_L ({\cal N}^\dagger  {\cal R})_{ij} N_{m_{j}} 
  + \rm{H.c.} \Big\} 
\Big] , 
\label{NC}
\eea
where $\theta_{\rm W}$ is the weak mixing angle.
Through the mixing ${\cal R}_{\alpha i}$, the heavy neutrinos can be produced at high energy colliders, 
   which have been extensively studied \cite{Das:2012ze, Dev:2013wba,  Das:2014jxa, Alva:2014gxa, Das:2015toa, Hessler:2014ssa, Degrande:2016aje, Das:2016hof,Das:2017pvt,  Antusch:2014woa, Antusch:2015mia,
Antusch:2015rma, Antusch:2015gjw, Antusch:2016brq, Antusch:2016vyf, Fischer:2016rsh, Antusch:2016qby, Antusch:2016ejd, Dib:2015oka, Dib:2016wge,Asaka:2013jfa, 
Dib:2014iga, Dib:2014pga, Cvetic:2012hd, Cvetic:2010rw, Cvetic:2016fbv, Zamora-Saa:2016ito,Rasmussen:2016njh, Bambhaniya:2014hla, Bambhaniya:2014kga,Bambhaniya:2016rbb,Blennow:2016jkn,Caputo:2016ojx,Deppisch:2015qwa,Dev:2014xea}.  
For example, the production cross section of the $i$-th generation heavy neutrino 
  at the Large Hadron Collider (LHC) 
  through the process $q \bar{q}^\prime \to \ell N_i$  
 ($u \bar{d} \to \ell_\alpha^{+} N_i$ 
  and $ {\bar u} d \to \ell_\alpha^{-} \overline{N_i}$) is given by 
\bea 
 \sigma(q \bar{q}^\prime \to \ell_\alpha N_i)
 = \sigma_{LHC} |{\cal R}_{\alpha i}|^2,  
\label{X_LHC} 
\eea
where $\sigma_{LHC}$ is the production cross section of the SM neutrino 
  when its mass is set to be $m_N^{~i}$. 
Similarly, the production cross section at an $e^+ e^-$ collider 
 such as the Large Electron-Positron Collider (LEP) and the International Linear Collider (ILC) 
 is given by 
\bea 
 \sigma(e^+ e^- \to \overline{\nu_\alpha} N_i)
 = \sigma_{LC} |{\cal R}_{\alpha i}|^2,  
\label{X_LC} 
\eea
where $\sigma_{LC}$ is the production cross section of the SM neutrino 
  at an $e^+ e^-$ collider when its mass is set to be $m_N^{~i}$,  
  and we have used the approximation ${\cal N}^\dagger {\cal R} \simeq U_{MNS}^\dagger {\cal R}$ 
  for $|\epsilon_{\alpha \beta}| \ll 1$ as we will find in the following.

The elements of the matrices ${\cal N}$ and ${\cal R}$ are constrained by the experimental data.
In the following analysis, we adopt, for the current neutrino oscillation data,  
  $\sin^{2}2{\theta_{13}}=0.092$ \cite{Neut5} 
  along with the other oscillation data \cite{Neut3}: 
 $\sin^2 2\theta_{12}=0.87$, $\sin^2 2\theta_{23}=1.0$, 
 $\Delta m_{12}^2 = m_2^2-m_1^2 = 7.6 \times 10^{-5}$ eV$^2$, 
 and $\Delta m_{23}^2= |m_3^2-m_2^2|=2.4 \times 10^{-3}$ eV$^2$. 
The neutrino mixing matrix is given by 
\bea
U_{\rm{PMNS}} = \begin{pmatrix} C_{12} C_{13}&S_{12}C_{13}&S_{13}e^{i\delta}\\-S_{12}C_{23}-C_{12}S_{23}S_{13}e^{i\delta}&C_{12}C_{23}-S_{12}S_{23}S_{13}e^{i\delta}&S_{23} C_{13}\\ S_{12}C_{23}-C_{12}C_{23}S_{13}e^{i\delta}&-C_{12}S_{23}-S_{12}C_{23}S_{13}e^{i\delta}&C_{23}C_{13} \end{pmatrix} \begin{pmatrix} 1&0&0\\0&e^{i\rho}&0\\0&0&1\end{pmatrix}
\label{pmns}
\eea
where $C_{ij}=\cos\theta_{ij}$ and $S_{ij}=\sin\theta_{ij}$. 
We consider the Dirac $CP$-phase $(\delta)$ and the Majorana phase $(\rho)$ 
  as free parameters. 

The minimal seesaw scenario predicts one massless eigenstate. 
For the light neutrino mass spectrum, we consider both the normal hierarchy (NH) and the inverted hierarchy (IH).  
In the NH case, the diagonal mass matrix is given by 
\bea 
  D_{\rm{NH}} ={\rm diag}
  \left(0, \sqrt{\Delta m_{12}^2},
           \sqrt{\Delta m_{12}^2 + \Delta m_{23}^2} \right),  
\label{DNH}
\eea 
while in the IH case 
\bea 
  D_{\rm{IH}} ={\rm diag}
\left( \sqrt{\Delta m_{23}^2 - \Delta m_{12}^2}, 
 \sqrt{\Delta m_{23}^2}, 0 \right).  
\label{DIH}
\eea 
In order to make our discussion simple, 
  we assume the degeneracy of the heavy neutrinos in mass such as $M_N= m_N^{~1}=m_N^{~2}$, 
 so that the light neutrino mass matrix is simplified as 
\bea 
   m_\nu = \frac{1}{M_N} m_D m_D^T 
 = U_{\rm{MNS}}^* D_{\rm{NH/IH}} U_{\rm{MNS}}^\dagger ,  
\eea  
for the NH/IH cases. 
 From this formula, we can parameterize the neutrino Dirac mass matrix as ~\cite{CI_Para}\footnote{
This formula only holds at the tree level and a generalization at the one-loop level has been introduced 
  in Ref.~\cite{Lopez-Pavon:2015cga}.  
Although the loop corrections can be potentially important in our analysis, 
   the loop corrections vanish when the heavy neutrinos are degenerate~\cite{Lopez-Pavon:2015cga},  
   and our analysis is reliable at the tree level.  }  
\bea 
  m_D = \sqrt{M_N} U_{\rm{MNS}}^* \sqrt{D_{\rm{NH/IH}}} \; O ,  
\label{mD}
\eea
where the matrices denoted as $\sqrt{D_{\rm{NH/IH}}}$ are defined as 
\bea
\sqrt{D_{\rm{NH}}} =
\begin{pmatrix}
    0 & 0 \\
 (\Delta m_{12}^2)^{\frac{1}{4}} & 0 \\
             0   & (\Delta m_{23}^2+\Delta m_{12}^2)^{\frac{1}{4}} \\
\end{pmatrix},  \; \; 
\sqrt{D_{\rm{IH}}} =
\begin{pmatrix}

    (\Delta m_{23}^2 - \Delta m_{12}^2)^{\frac{1}{4}} & 0 \\
    0 & (\Delta m_{23}^2)^{\frac{1}{4}} \\
    0 & 0 \\
\end{pmatrix}, 
\eea 
and 
$O$ is a general $2\times2$ orthogonal matrix given by
\bea
O=\begin{pmatrix}
\cos(X+i Y)&\sin(X+i  Y)\\
-\sin(X+i Y)&\cos(X+i Y)
\end{pmatrix}
=\begin{pmatrix}
\cosh Y &i \sinh Y\\
-i\sinh Y & \cosh Y
\end{pmatrix}
\begin{pmatrix}
\cos X & \sin  X \\
-\sin X & \cos X
\end{pmatrix}, 
\eea
where $X$ and $Y$ are real parameters.

Due to its non-unitarity, 
 the elements of the mixing matrix ${\cal N}$ 
 are severely constrained by the combined data 
 from the neutrino oscillation experiments, 
 the precision measurements of weak boson decays, 
 and the lepton-flavor-violating decays 
 of charged leptons~\cite{Constraints1, Constraints2, Constraints3, Constraints4, Constraints5}. 
We update the results by using more recent data 
 on the lepton-favor-violating decays~\cite{Adam, Aubert, OLeary}: 
\bea
|{\cal N}{\cal N}^\dagger| =
\begin{pmatrix} 
 0.994\pm0.00625& <1.288 \times 10^{-5} &  < 8.76356\times 10^{-3}\\
 <1.288 \times 10^{-5} & 0.995\pm 0.00625 & <1.046\times 10^{-2}\\
 < 8.76356 \times 10^{-3}& < 1.046 \times 10^{-2} & 0.995\pm 0.00625
\end{pmatrix} , 
\eea
where the diagonal elements are from the precision measurements of weak boson decays 
 (the SM prediction is $1$)  
  while the off-diagonal elements are the upper bounds from the lepton-favor-violating decays, 
  namely, the (1,2) and (2,1) elements from the $\mu \to e \gamma$ process, 
  the (2,3) and (3,2) elements from the $\tau \to \mu \gamma$ process, 
  and the (1,3) and (3,1) elements from the $\tau \to e \gamma$ process.   
Since ${\cal N}{\cal N}^\dagger \simeq {\bf 1} - \epsilon$, 
 we have the constraints on $\epsilon$ such that 
\bea
|\epsilon| =
\begin{pmatrix} 
 0.006\pm0.00625& < 1.288 \times 10^{-5} & < 8.76356 \times 10^{-3}\\
 < 1.288\times 10^{-5} & 0.005\pm 0.00625 & < 1.046 \times 10^{-2}\\
 < 8.76356\times 10^{-3}& < 1.046 \times 10^{-2} & 0.005\pm 0.00625
\end{pmatrix} .
\label{eps}
\eea
The most stringent bound is given by the $(1,2)$-element 
 which is from the constraint on the lepton-flavor-violating 
 muon decay $\mu \to e \gamma$.
Using the general parametrization of the Dirac mass matrix in Eq.~(\ref{mD}), we have 
\bea
\epsilon(\delta,\rho,Y)
  &=&(\mathcal{R}^\ast \mathcal{R}^T)_{\rm{NH/IH}} = \frac{1}{M_N^2} m_D m_D^T \nonumber \\ 
    &=& \frac{1}{m_{N}}U_{\rm{MNS}} \sqrt{D_{\rm{NH/IH}}} O^{\ast} O^{T} \sqrt{D_{\rm{NH/IH}}} U_{\rm{MNS}}^{\dagger}. 
\label{RR}    
\eea 
Here, note that $\epsilon(\delta,\rho,Y)$ is independent of $X$ since 
\bea
O^\ast O^T=
\begin{pmatrix}
\cosh^2 Y + \sinh^2 Y & -2i \cosh Y \; \sinh Y \\
2i \cosh Y \; \sinh Y &\cosh^2 Y + \sinh^2 Y
\end{pmatrix}. 
\label{OO}
\eea
Now we perform a scan for the parameter set $\{\delta, \rho, Y\}$ 
   and identify an allowed region for which $\epsilon(\delta,\rho, Y)$ 
   satisfies the experimental constraints in Eq. (\ref{eps}).\footnote{  
Similar analysis of the parameter scan have been done in Refs.~\cite{Asaka:2011pb, Ruchayskiy:2011aa, Gorbunov:2014ypa, Drewes:2015iva,Hernandez:2016kel,Drewes:2016jae}, but for heavy Majorana neutrinos (much) lighter than the weak bosons.  
In this paper, we focus on the Majorana neutrinos heavier than the weak bosons 
  from the view point of the direct heavy neutrino production at the LHC. 
Our resultant upper bounds on the mixing parameters are quite different from those obtained in the previous work. 
}

 \begin{figure}
\begin{center}
\includegraphics[scale=0.4]{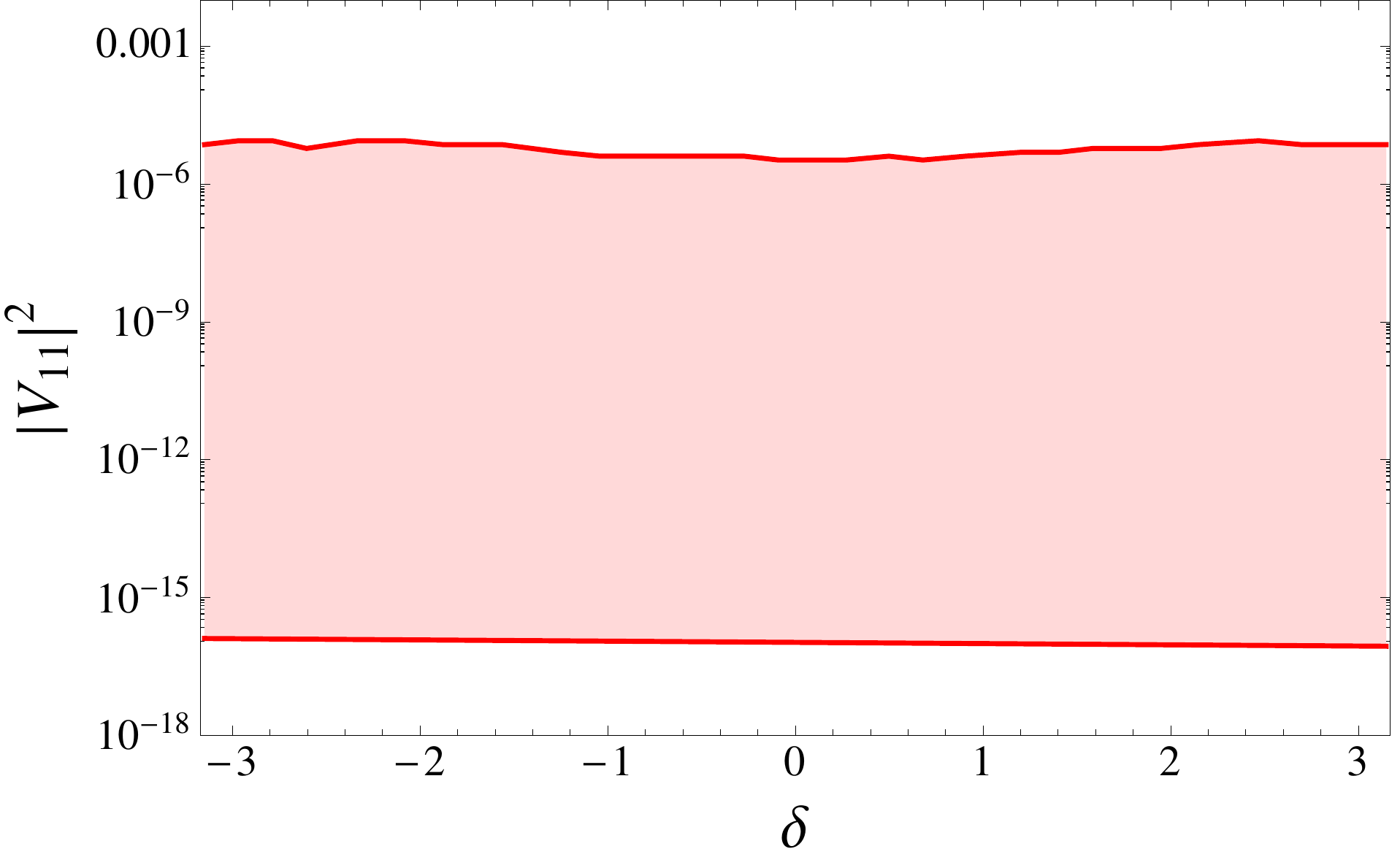}
\includegraphics[scale=0.4]{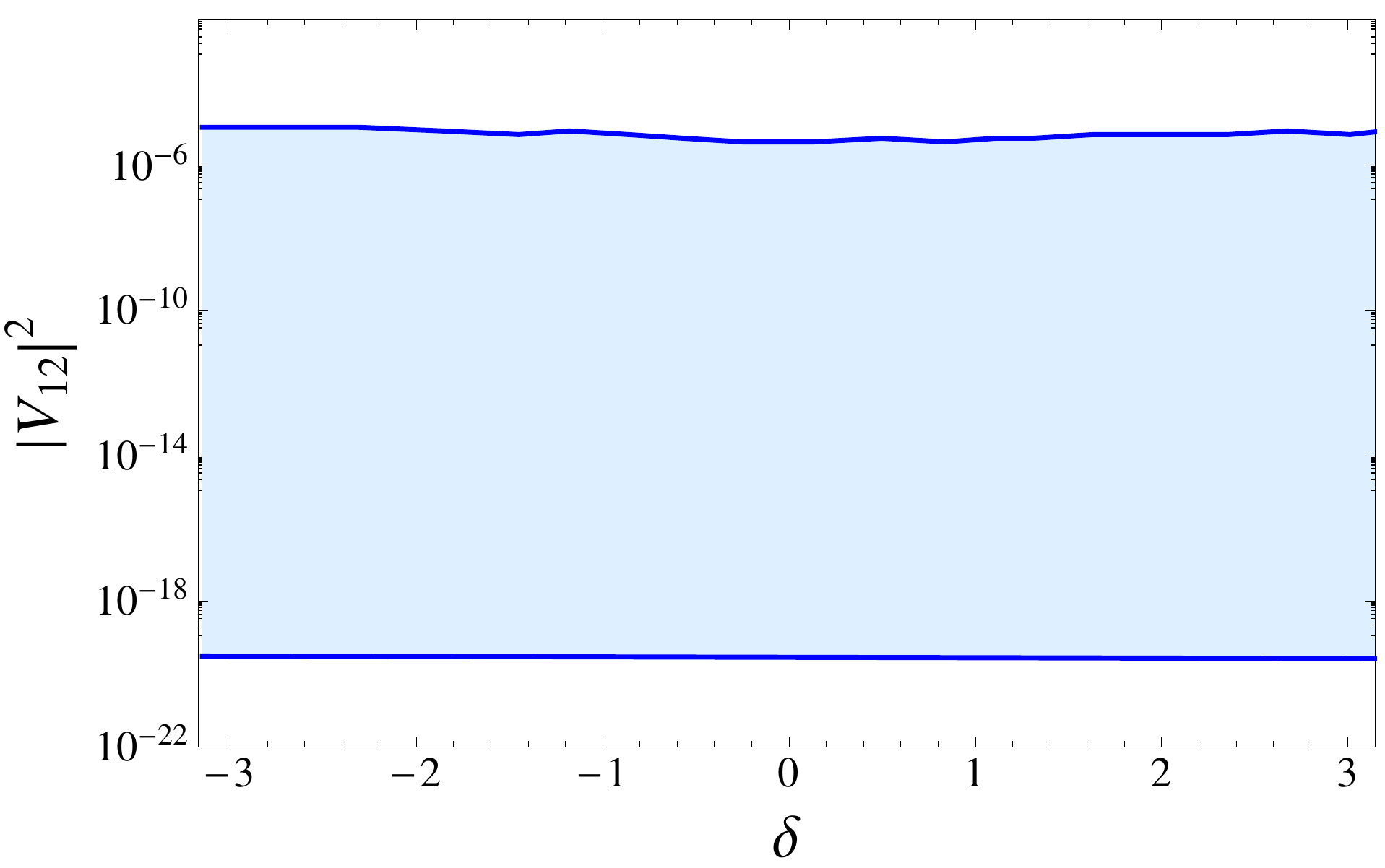}\\
\includegraphics[scale=0.4]{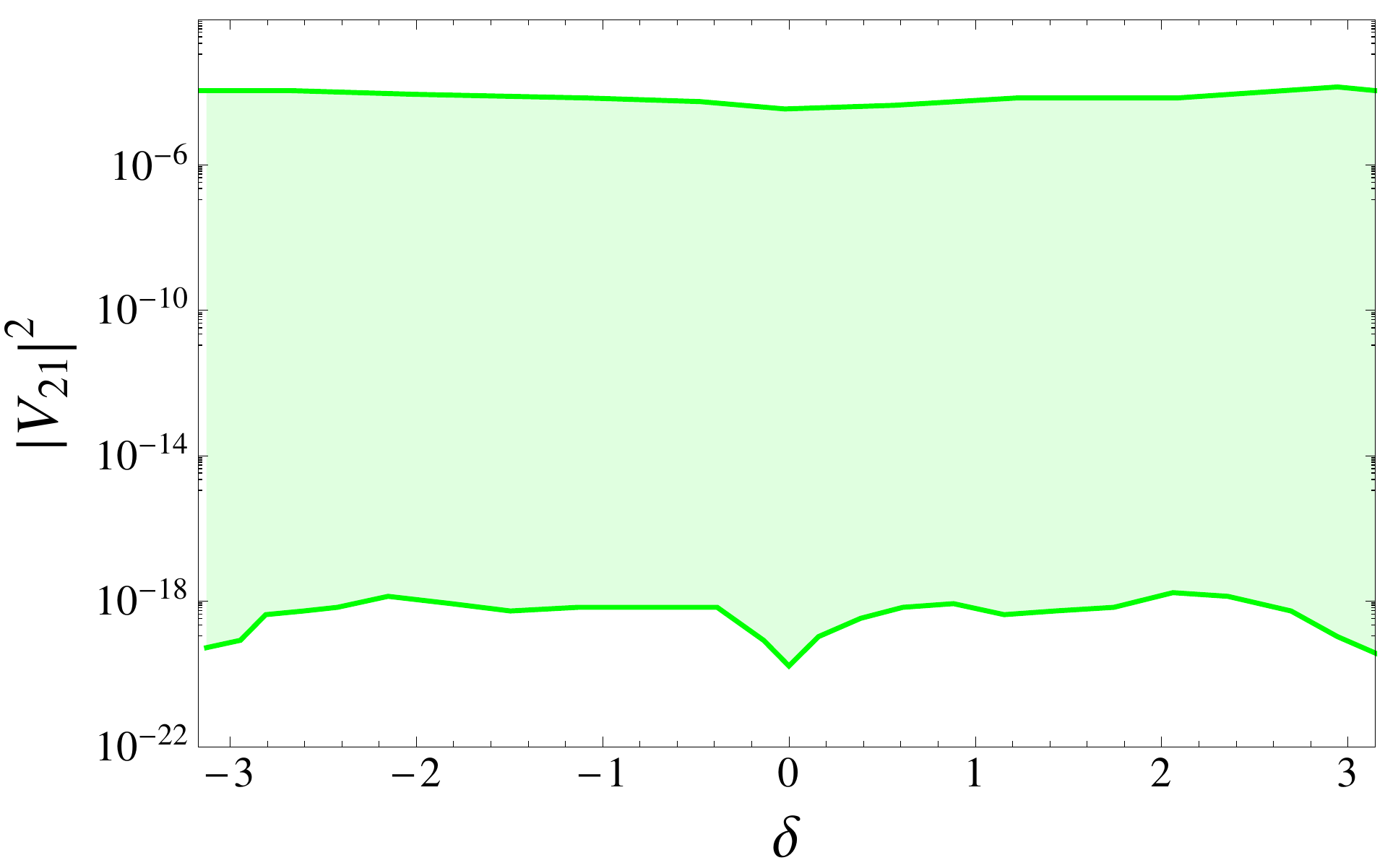}
\includegraphics[scale=0.4]{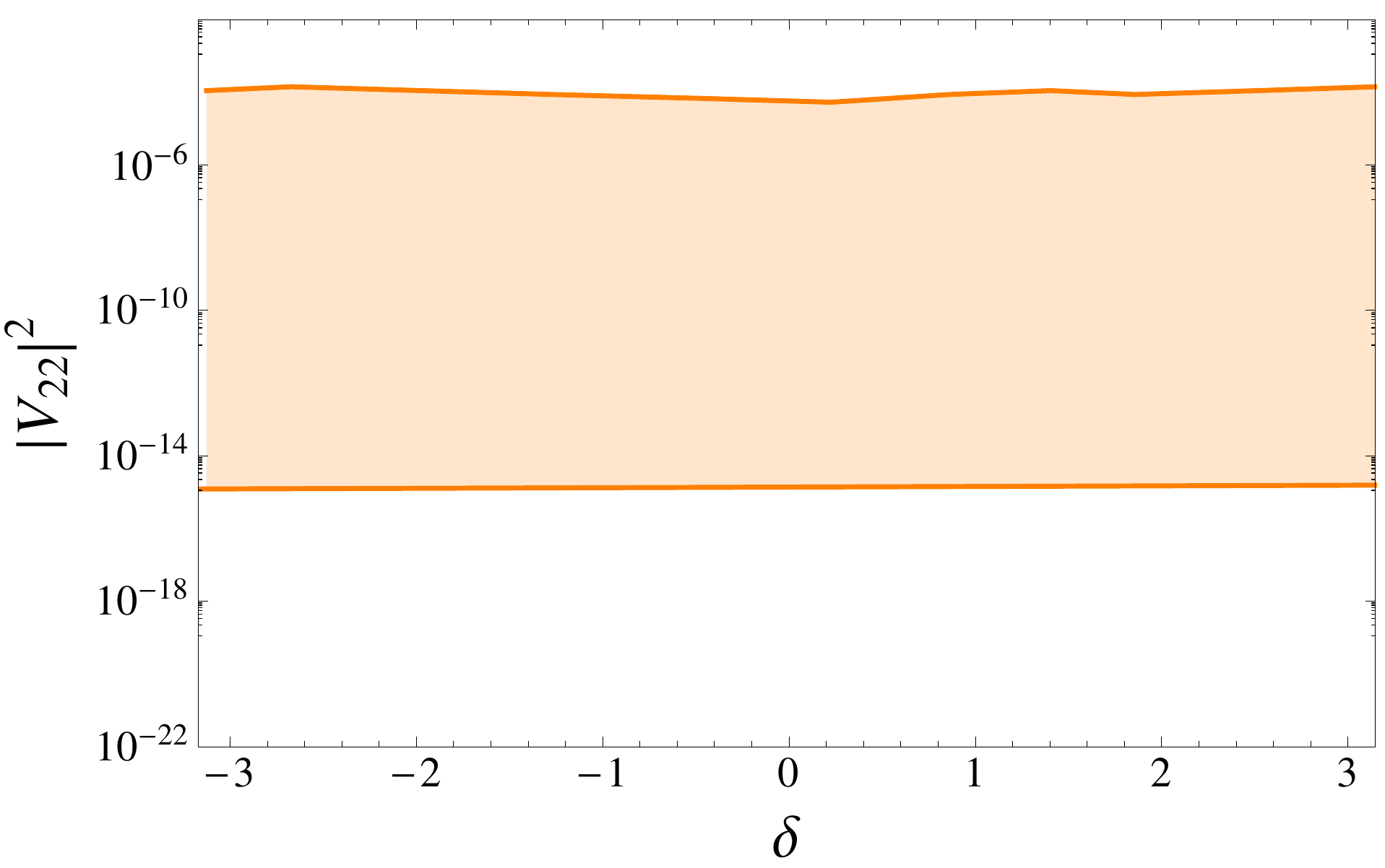}\\
\includegraphics[scale=0.4]{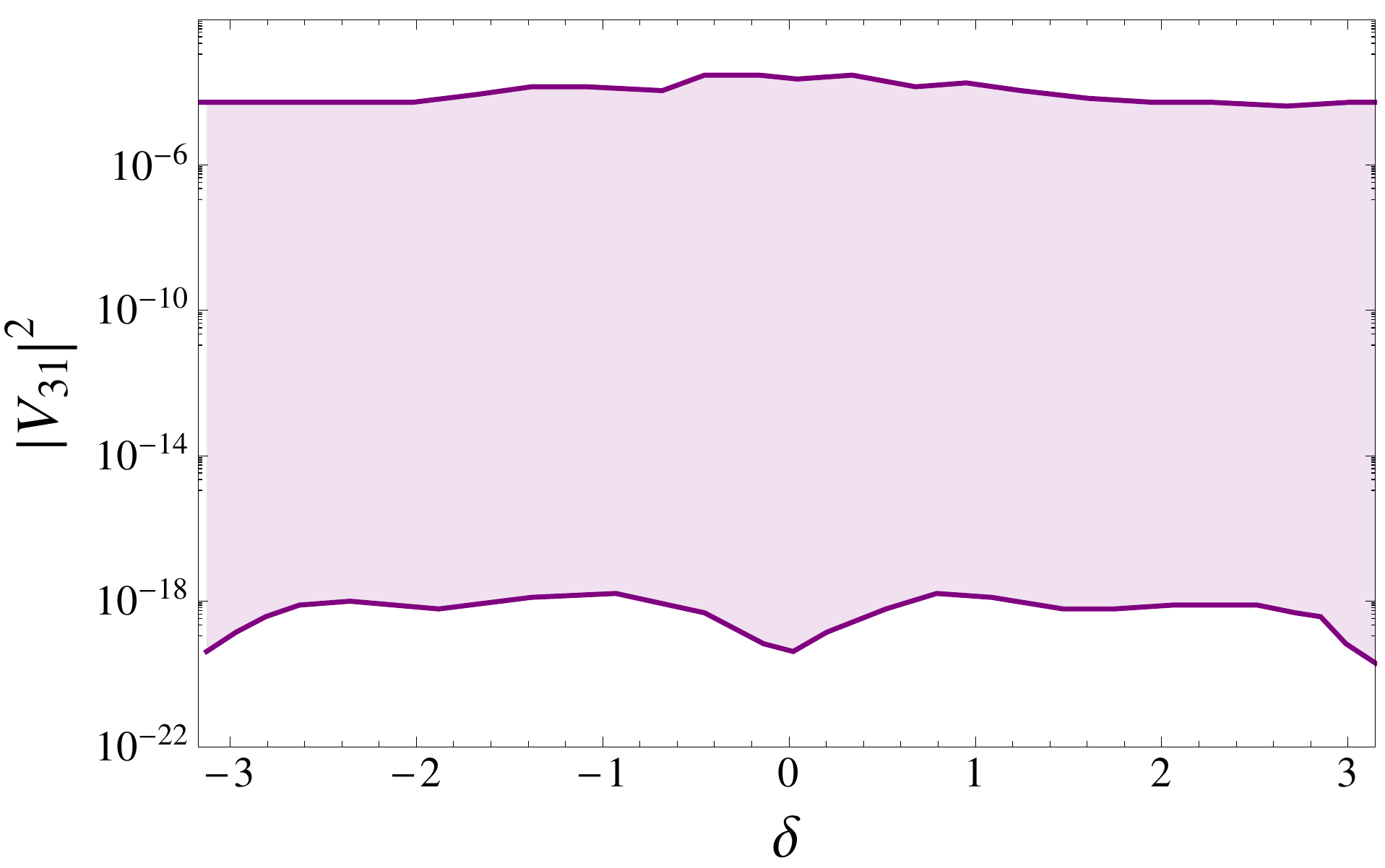}
\includegraphics[scale=0.4]{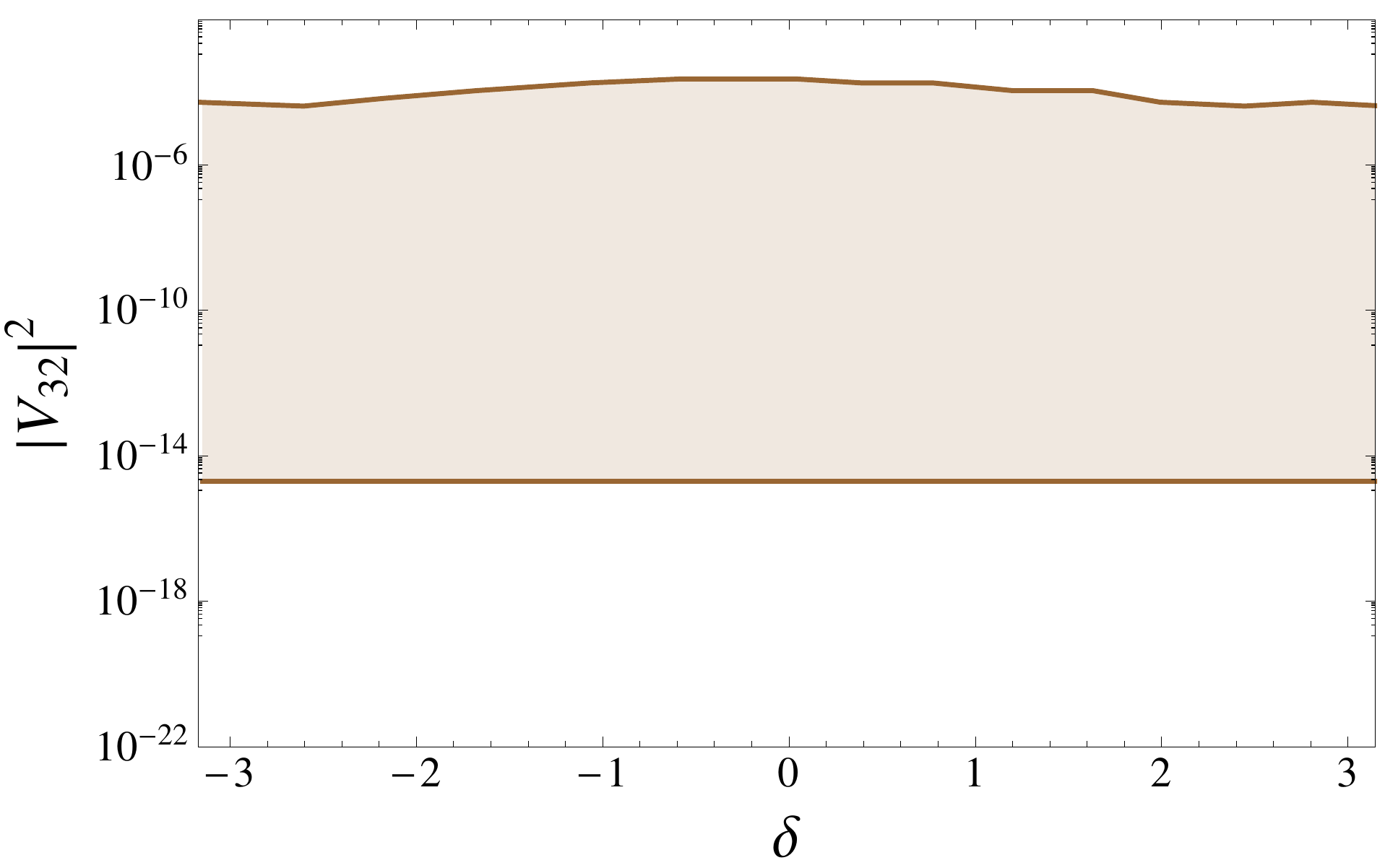}
\end{center}
\caption{
The experimental constraints on the mixing matrix elements $|{\cal R}_{\alpha i}|^2=|V_{\alpha i}|^2$ 
  in the NH case.  
The allowed region is shaded. 
The results are shown with respect to $ -\pi < \delta < \pi$. 
}
\label{mixNH1a}
\end{figure}
\begin{figure}
\begin{center}
\includegraphics[scale=0.4]{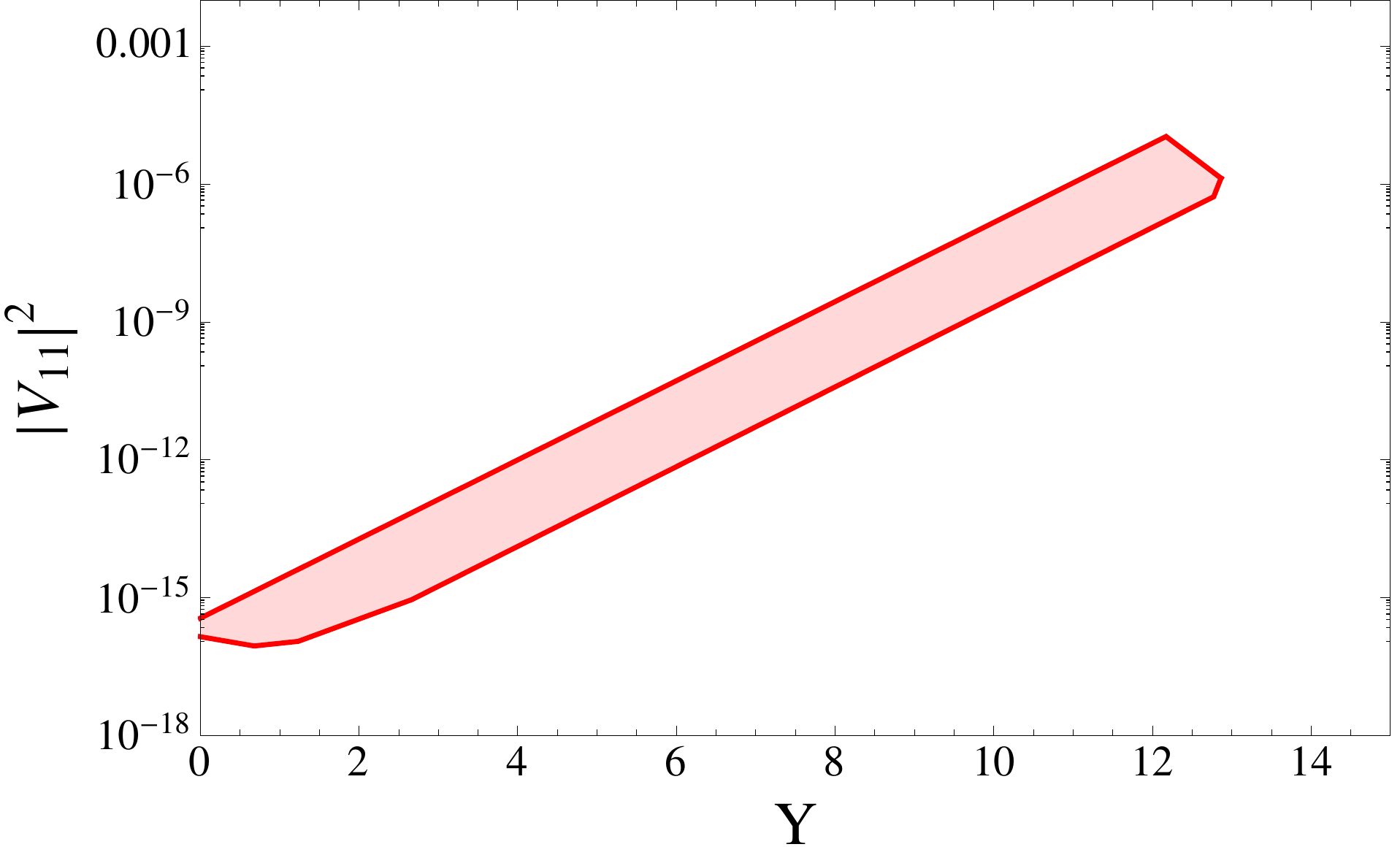}
\includegraphics[scale=0.4]{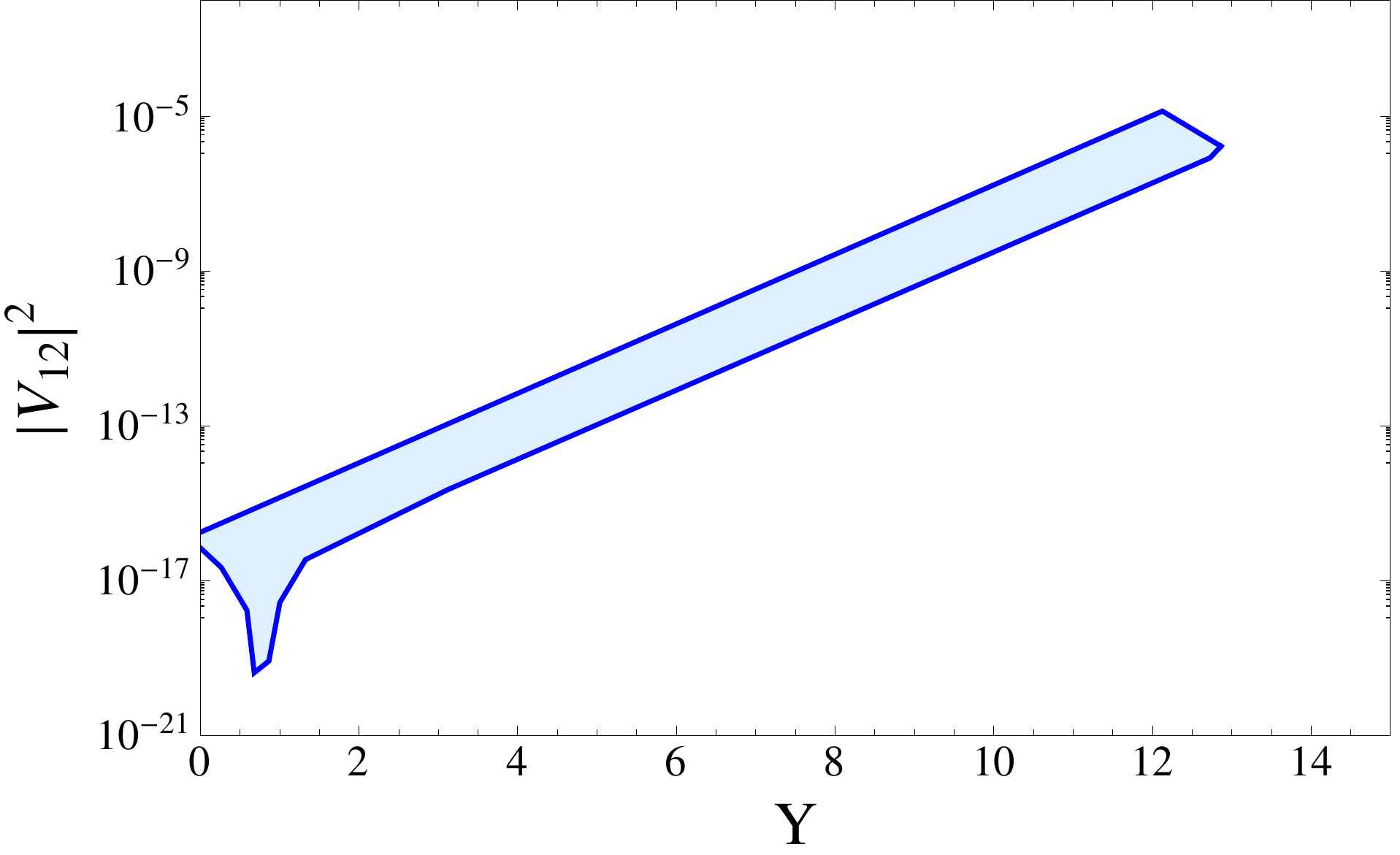}\\
\includegraphics[scale=0.4]{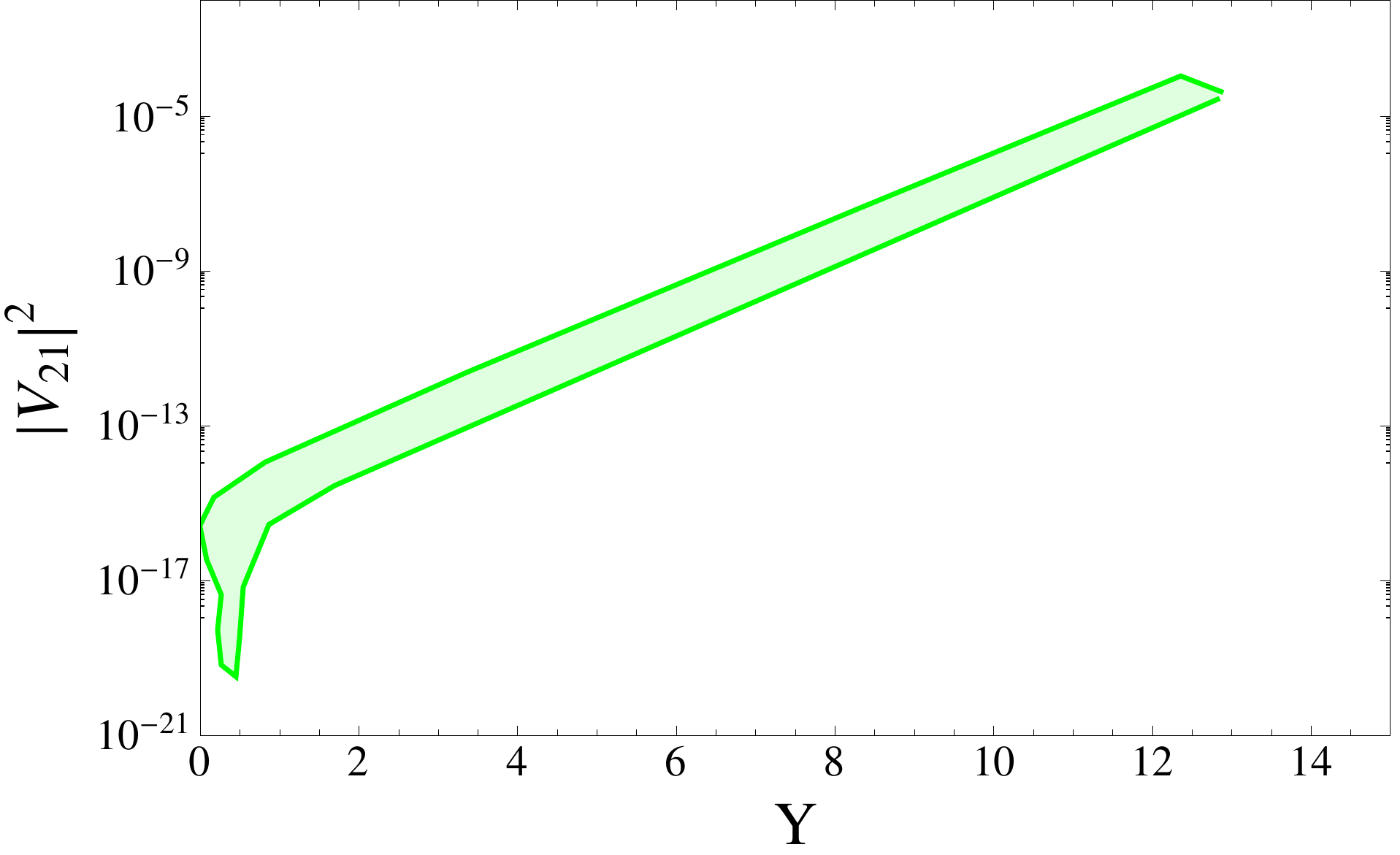}
\includegraphics[scale=0.4]{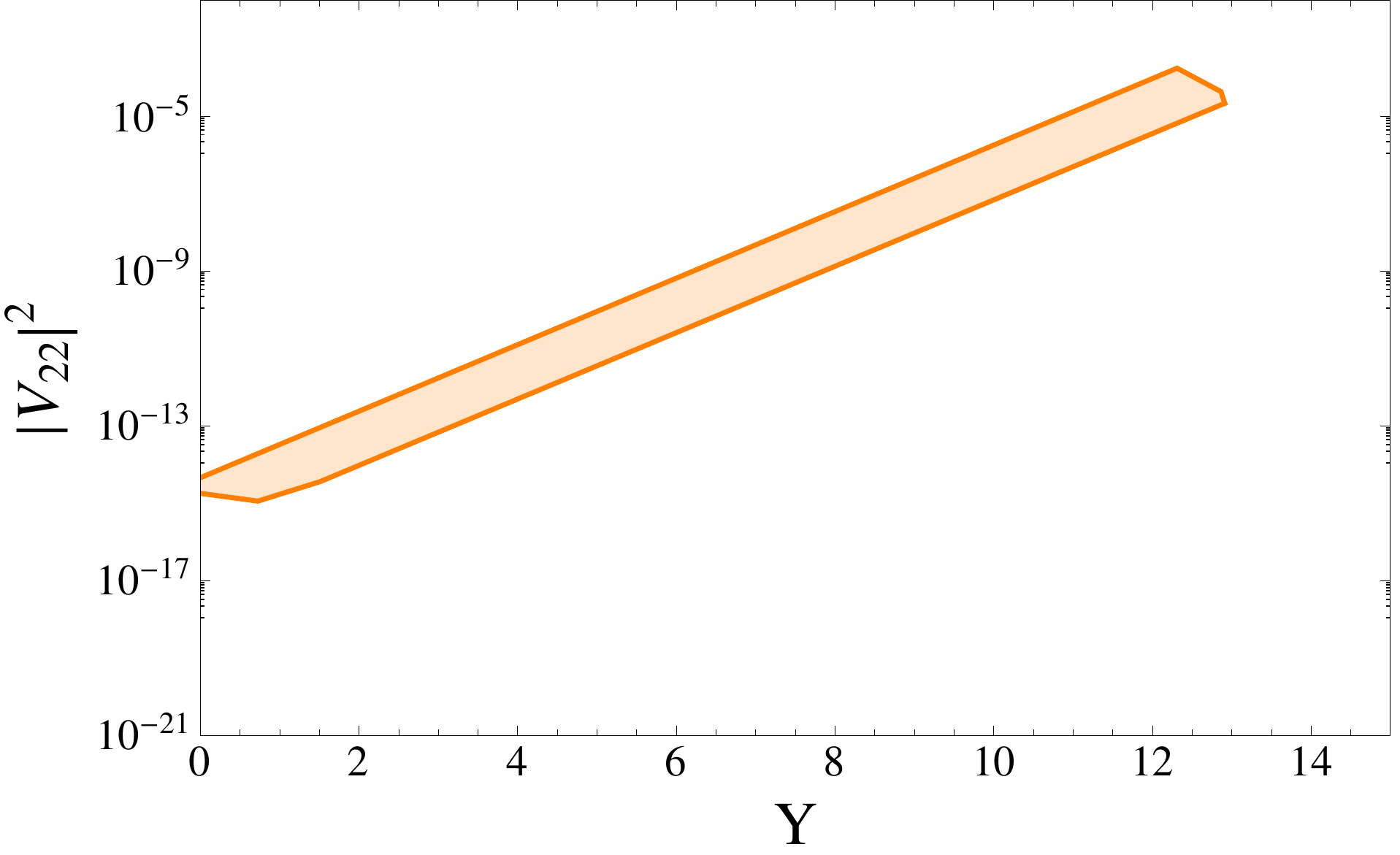}\\
\includegraphics[scale=0.4]{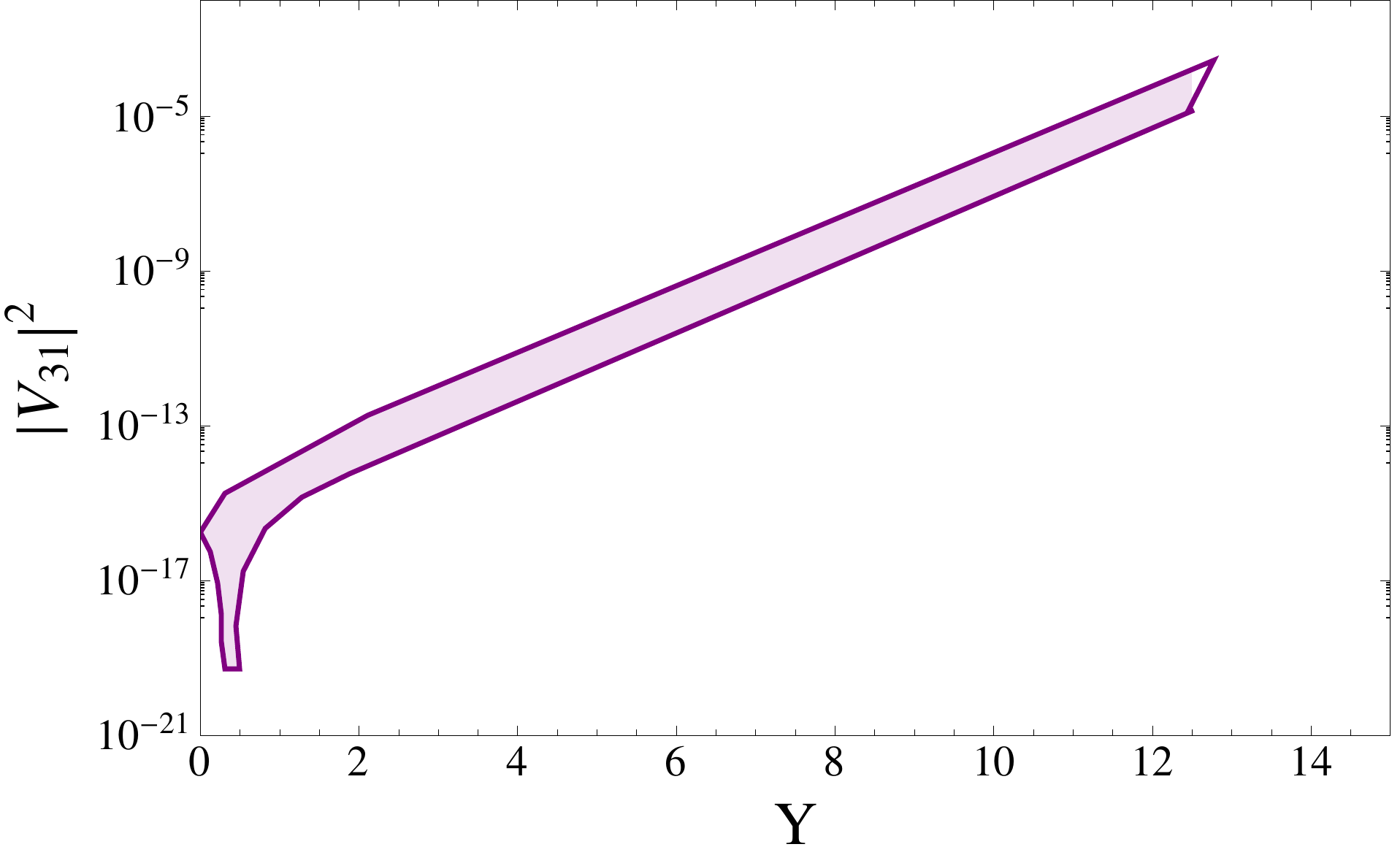}
\includegraphics[scale=0.4]{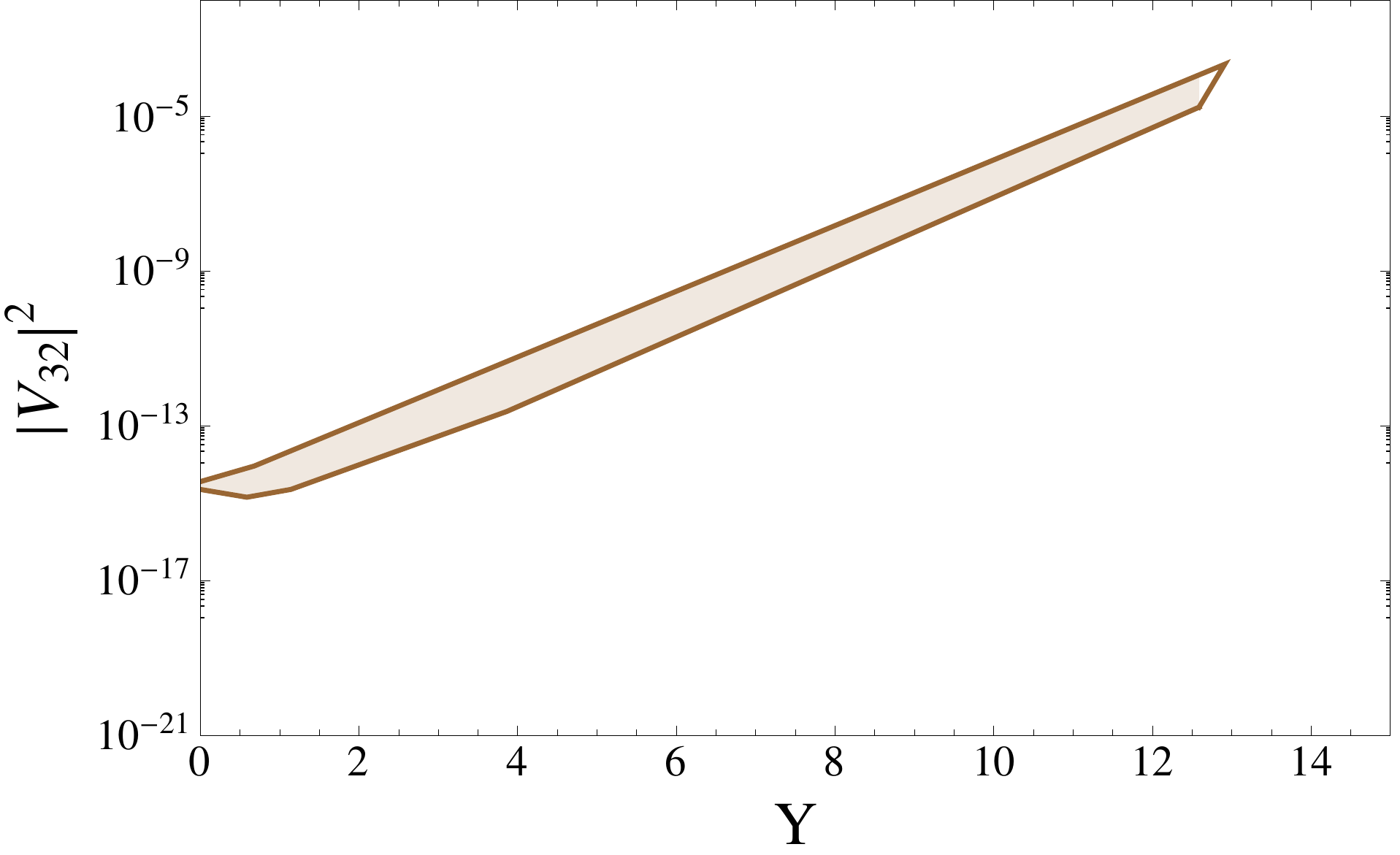}
\end{center}
\caption{
The experimental constraints on the mixing matrix elements $|{\cal R}_{\alpha i}|^2= |V_{\alpha i}|^2$ 
  in the NH case.  
The allowed region is shaded. 
The results are shown with respect to $Y$. 
}
\label{mixNH1b}
\end{figure}
 \begin{figure}
\begin{center}
\includegraphics[scale=0.4]{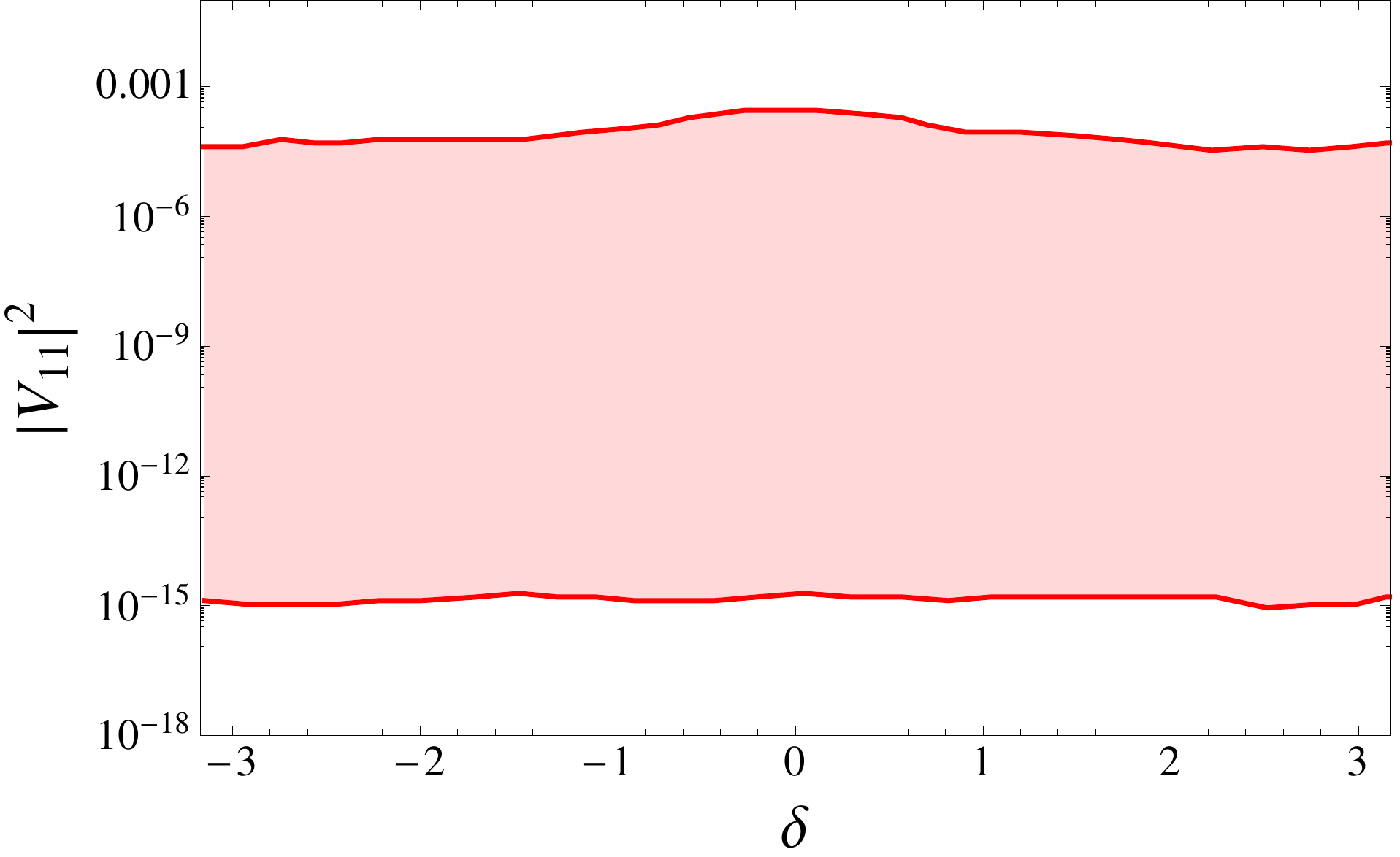}
\includegraphics[scale=0.4]{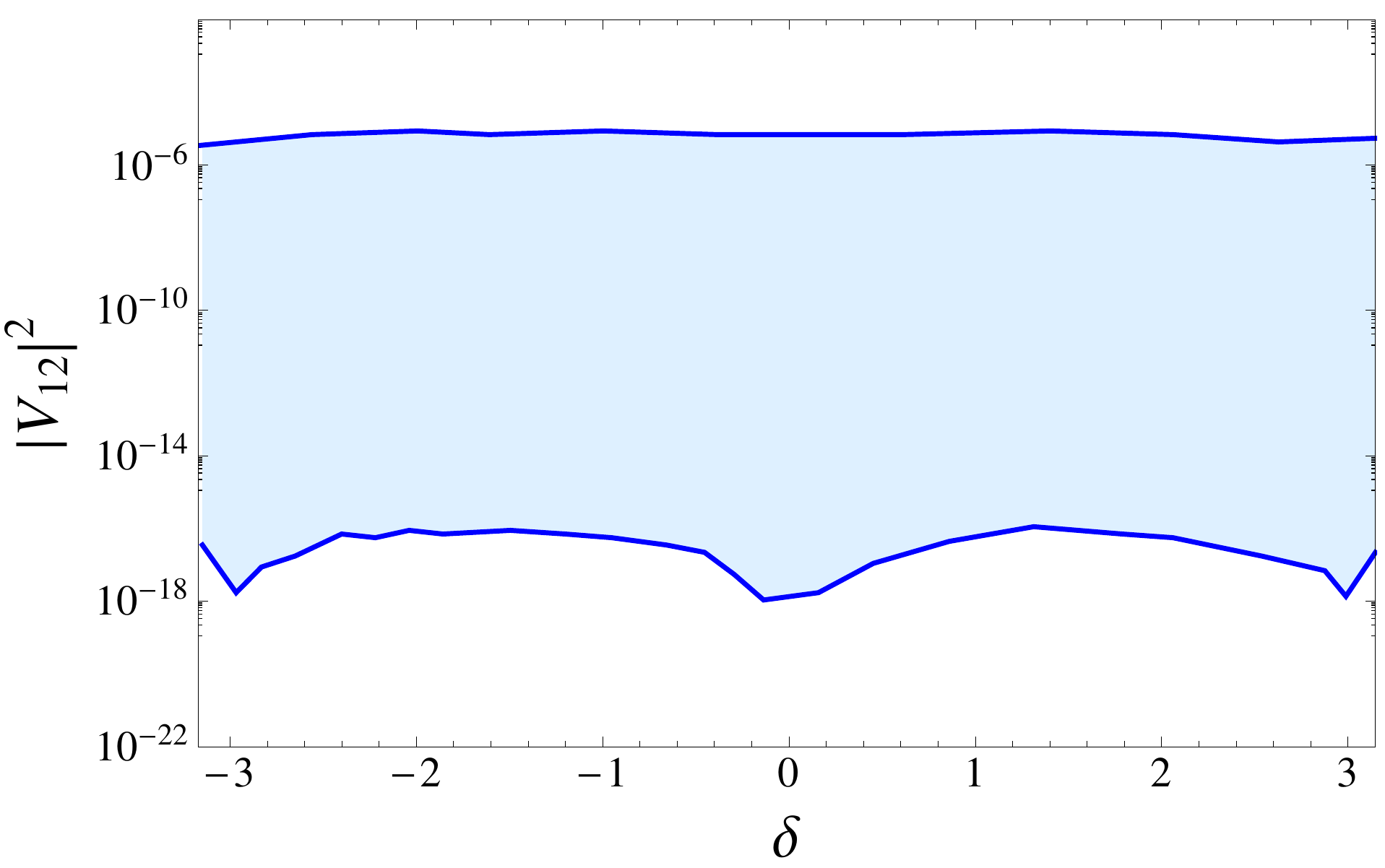}\\
\includegraphics[scale=0.4]{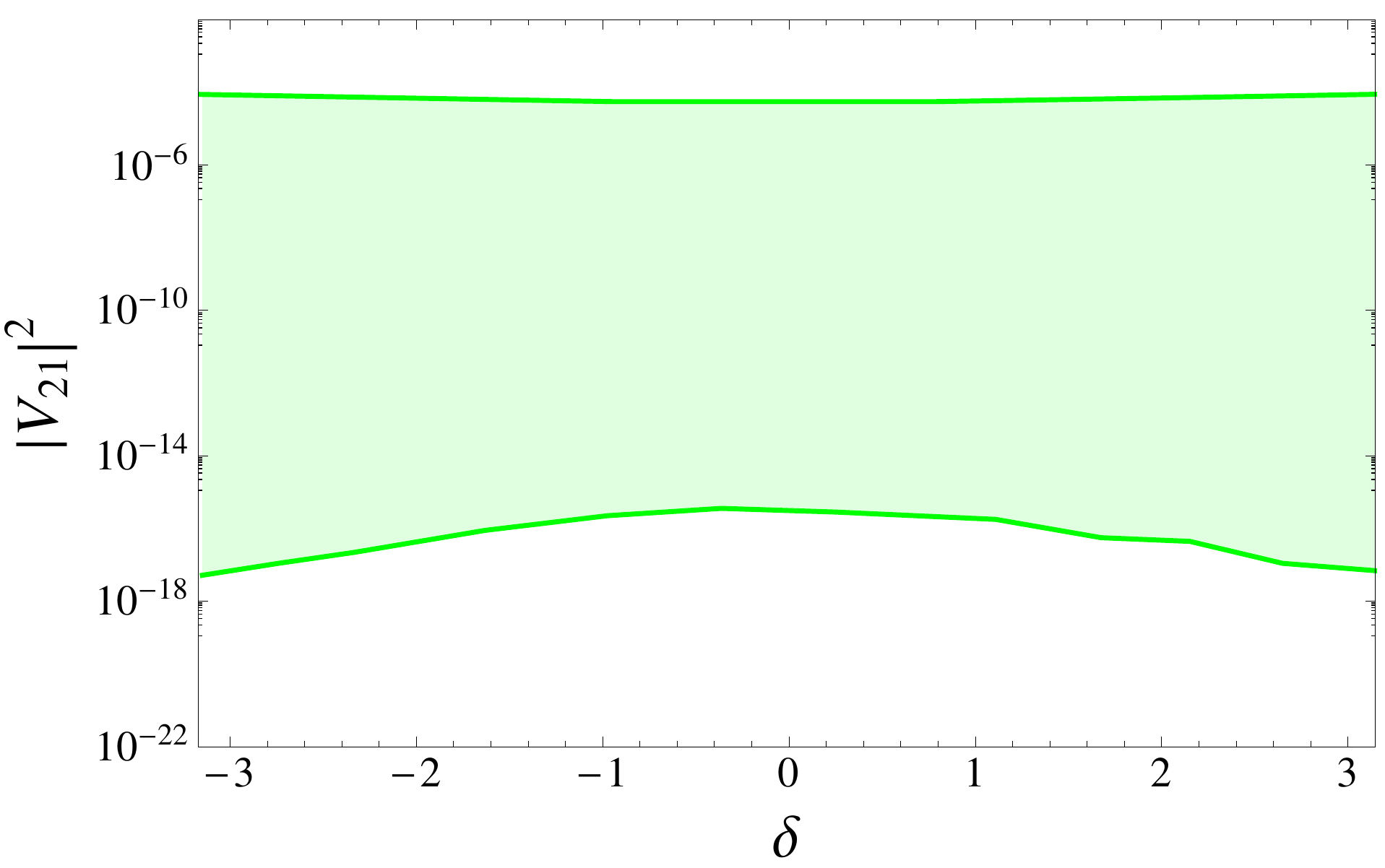}
\includegraphics[scale=0.4]{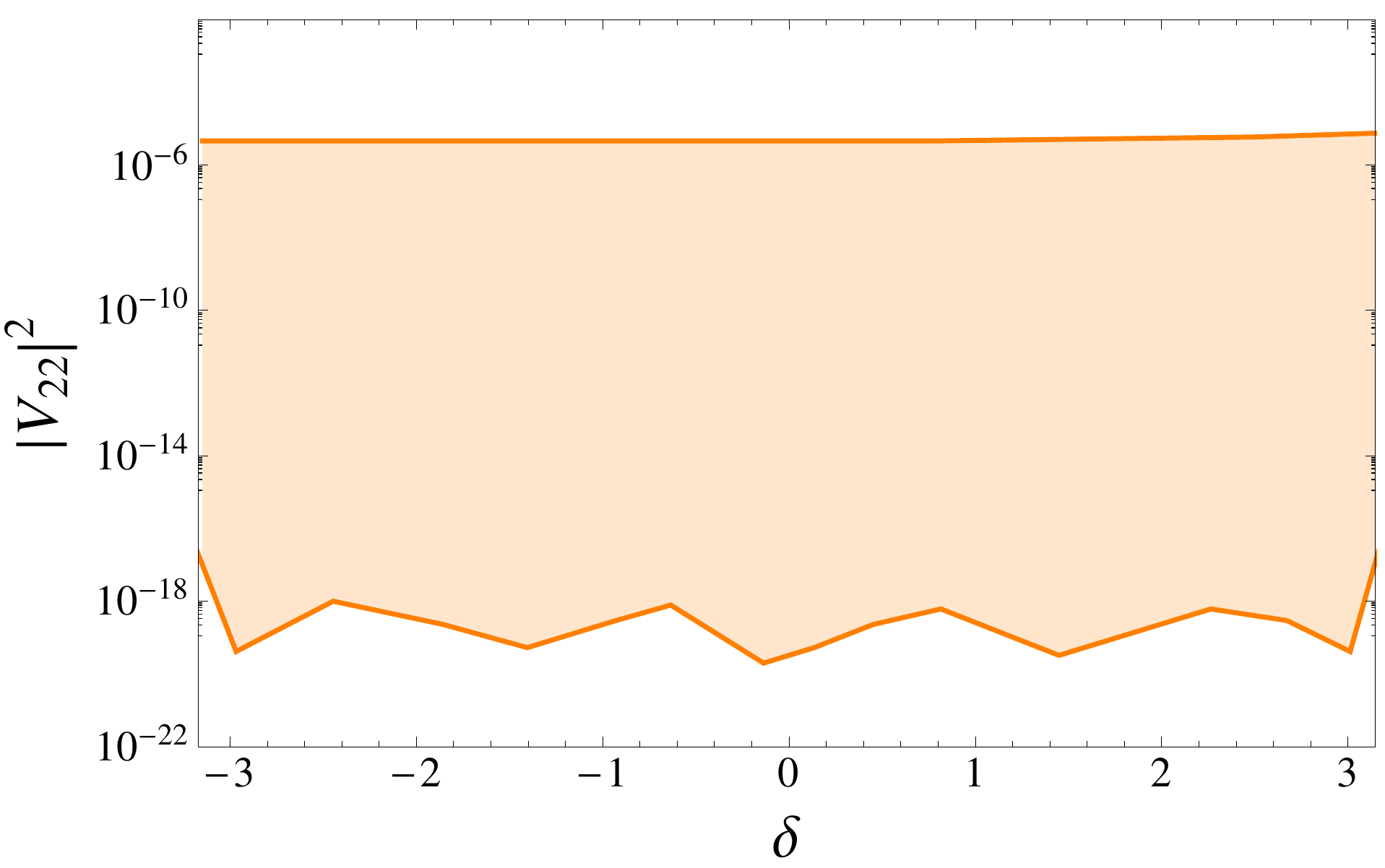}\\
\includegraphics[scale=0.4]{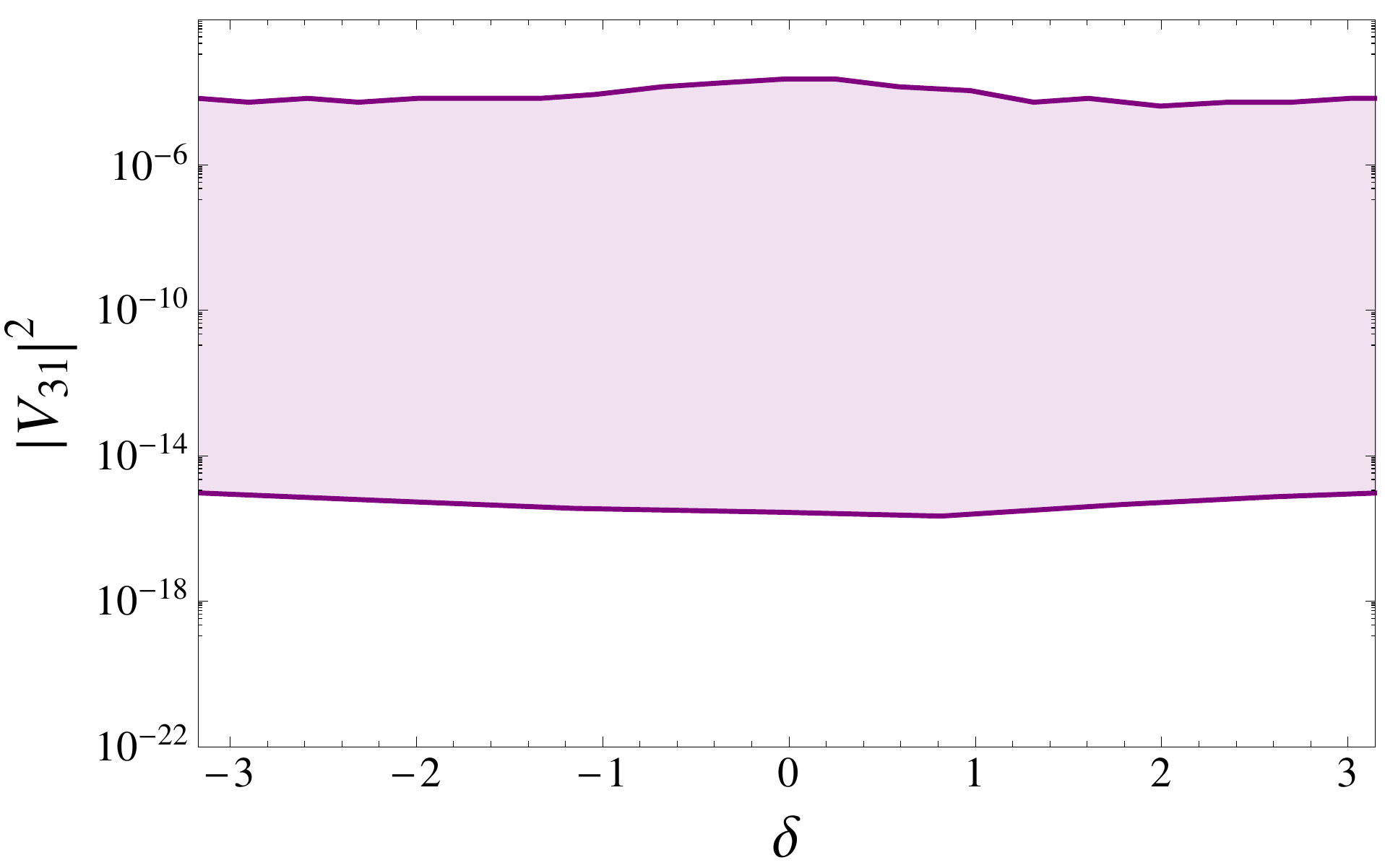}
\includegraphics[scale=0.4]{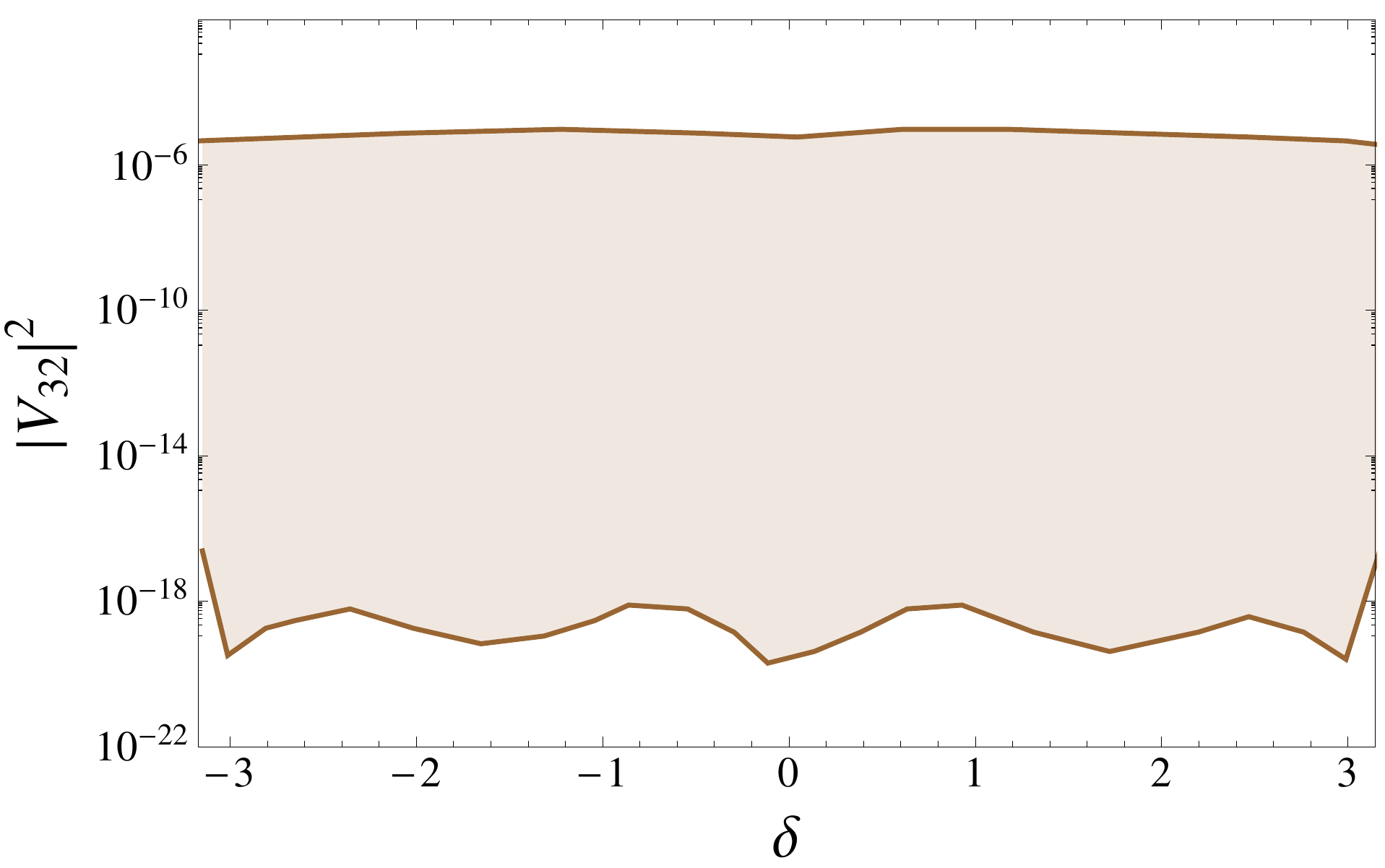}
\end{center}
\caption{
Same as Fig.~\ref{mixNH1a} but for the IH case. 
}
\label{mixIH1a}
\end{figure}
\begin{figure}
\begin{center}
\includegraphics[scale=0.4]{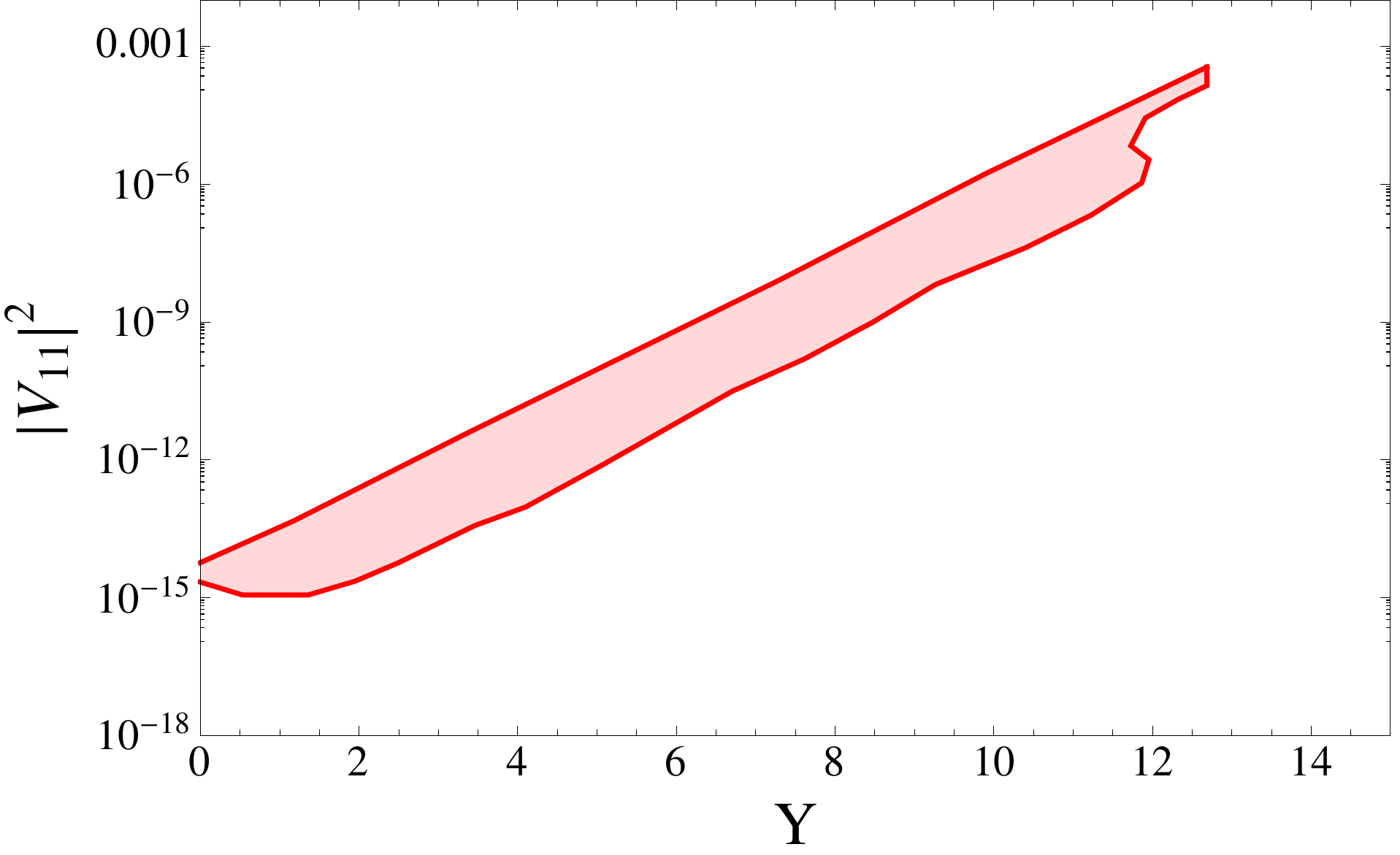}
\includegraphics[scale=0.4]{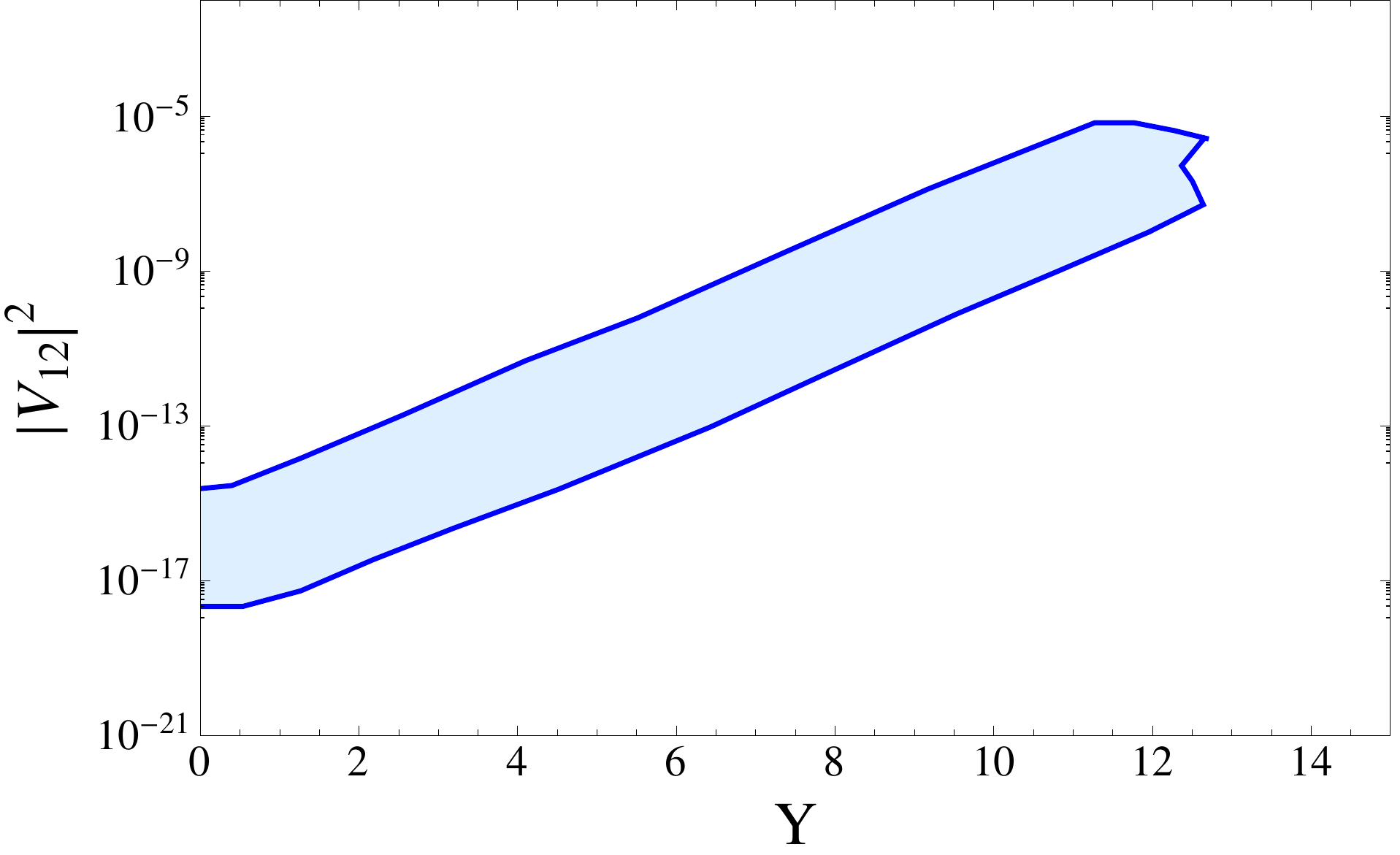}\\
\includegraphics[scale=0.4]{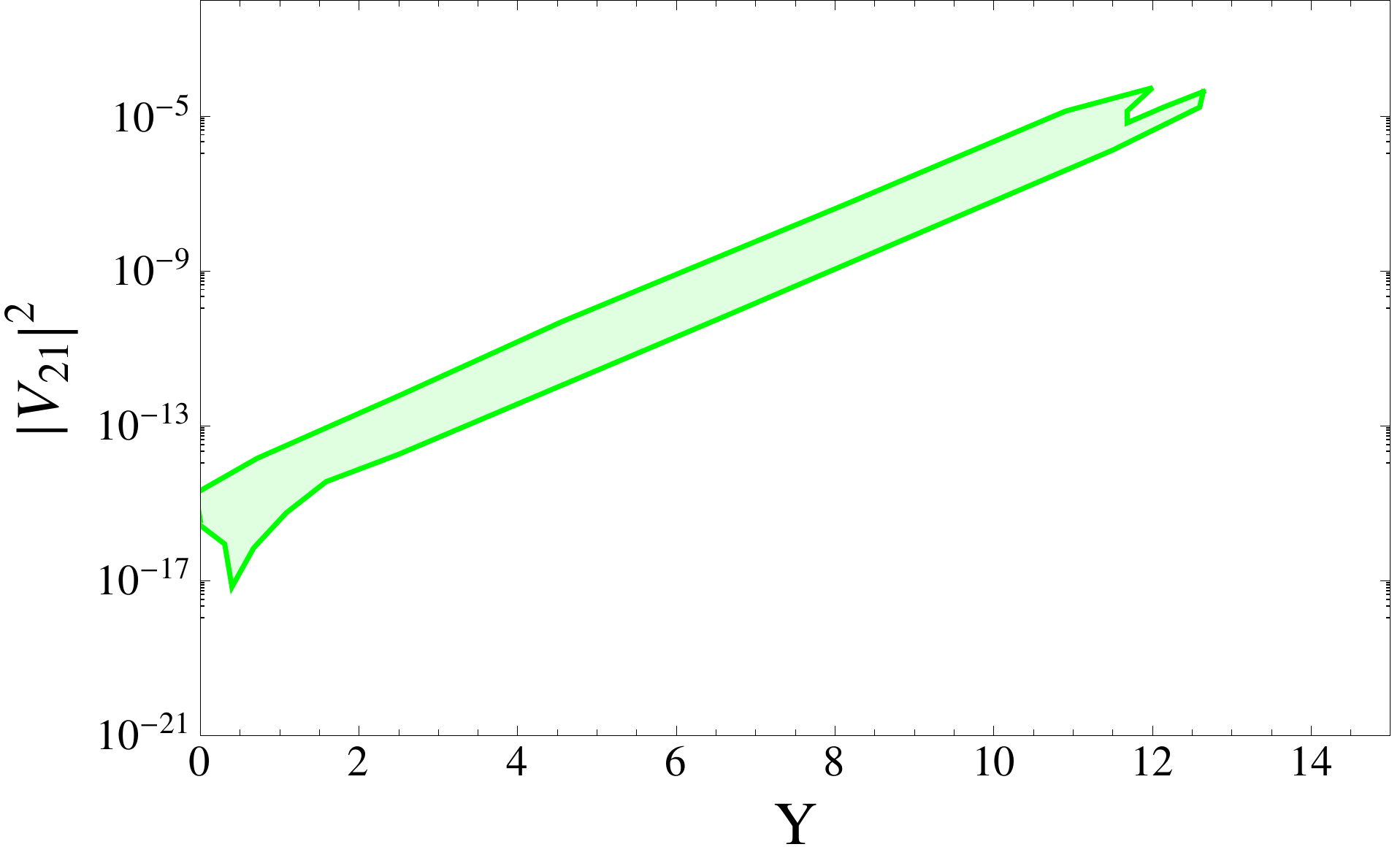}
\includegraphics[scale=0.4]{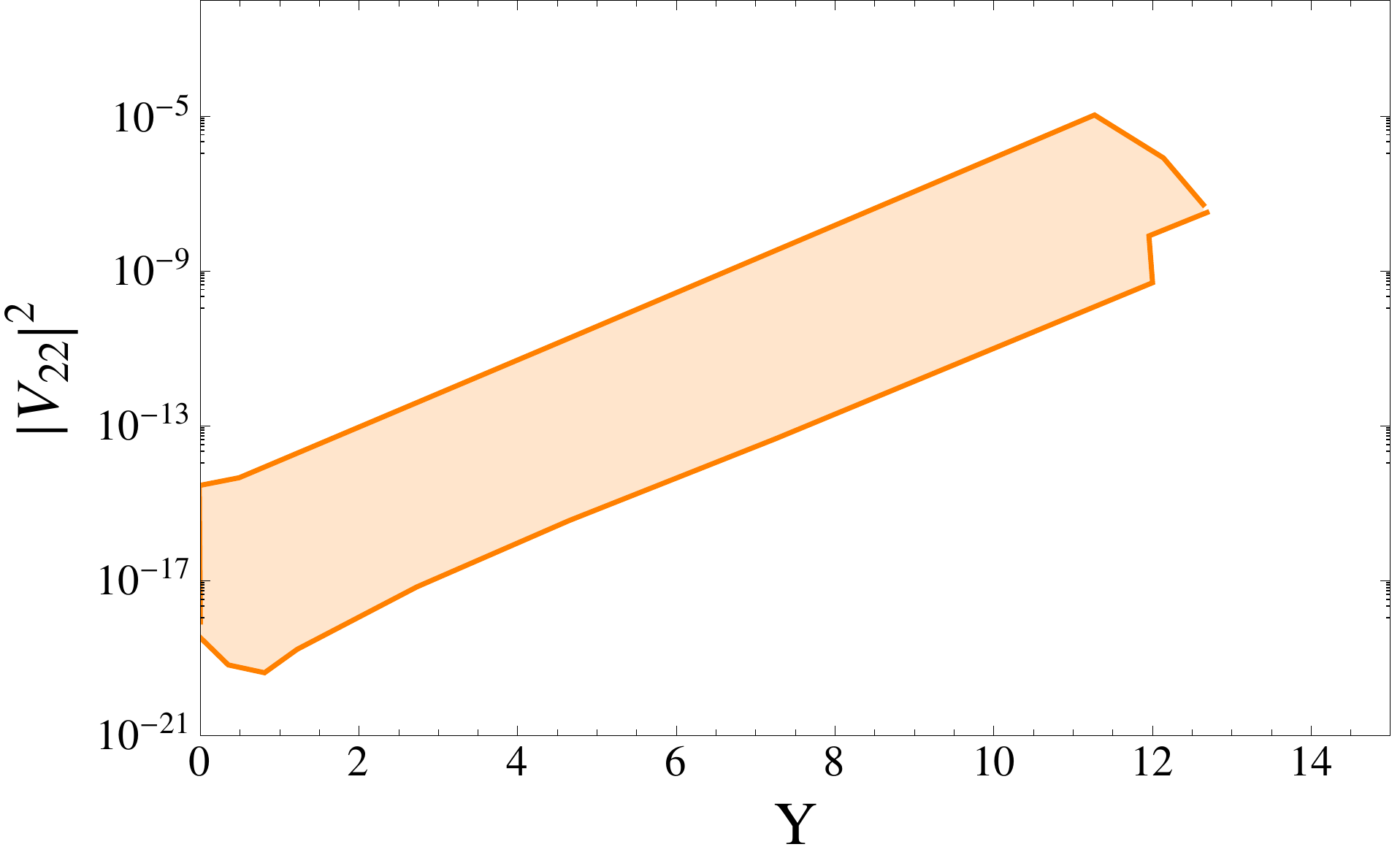}\\
\includegraphics[scale=0.4]{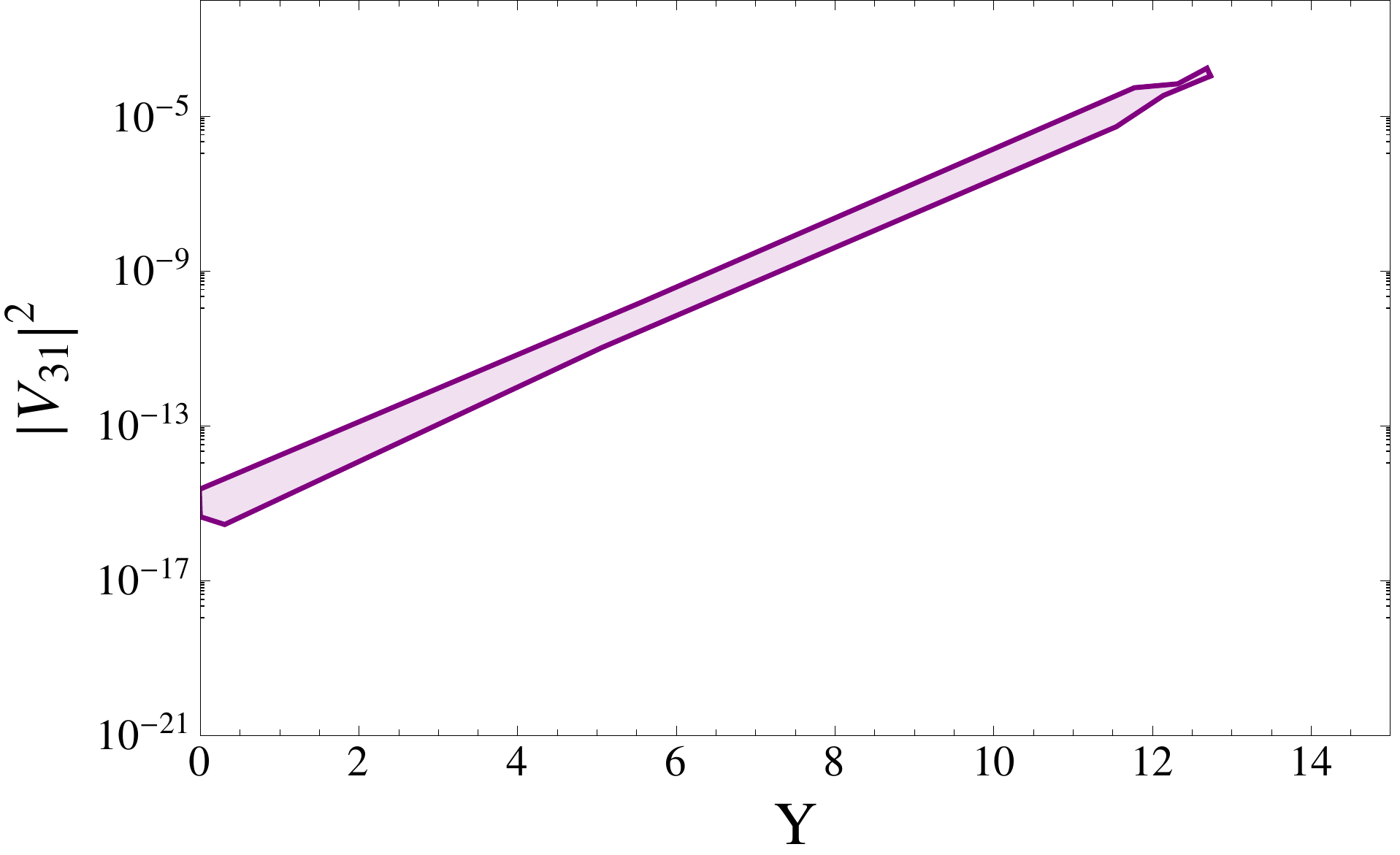}
\includegraphics[scale=0.4]{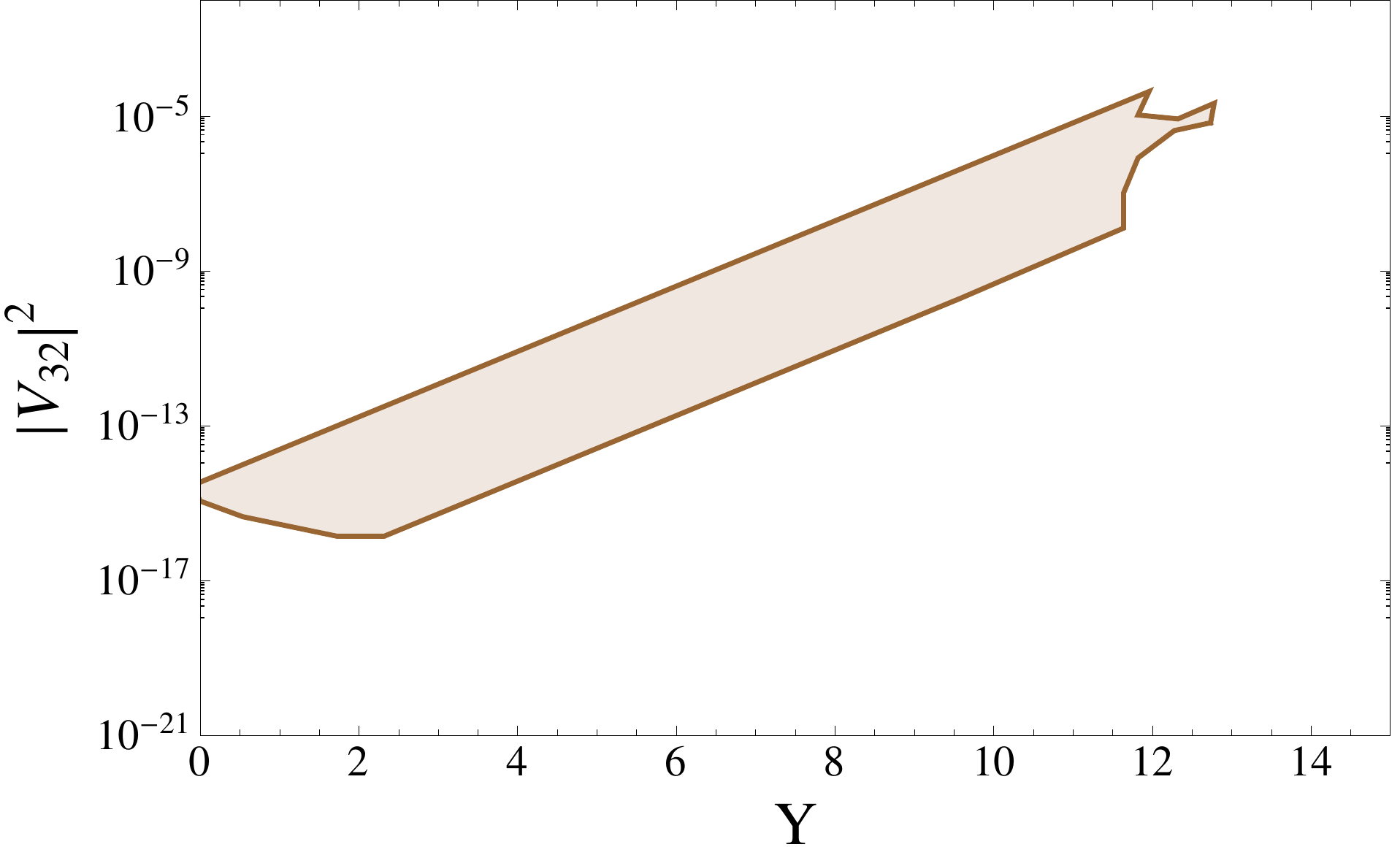}
\end{center}
\caption{
Same as Fig.~\ref{mixNH1b} but for the IH case. 
}
\label{mixIH1b}
\end{figure}

\begin{figure}
\begin{center}
\includegraphics[width=0.5\textwidth,angle=0,scale=0.9]{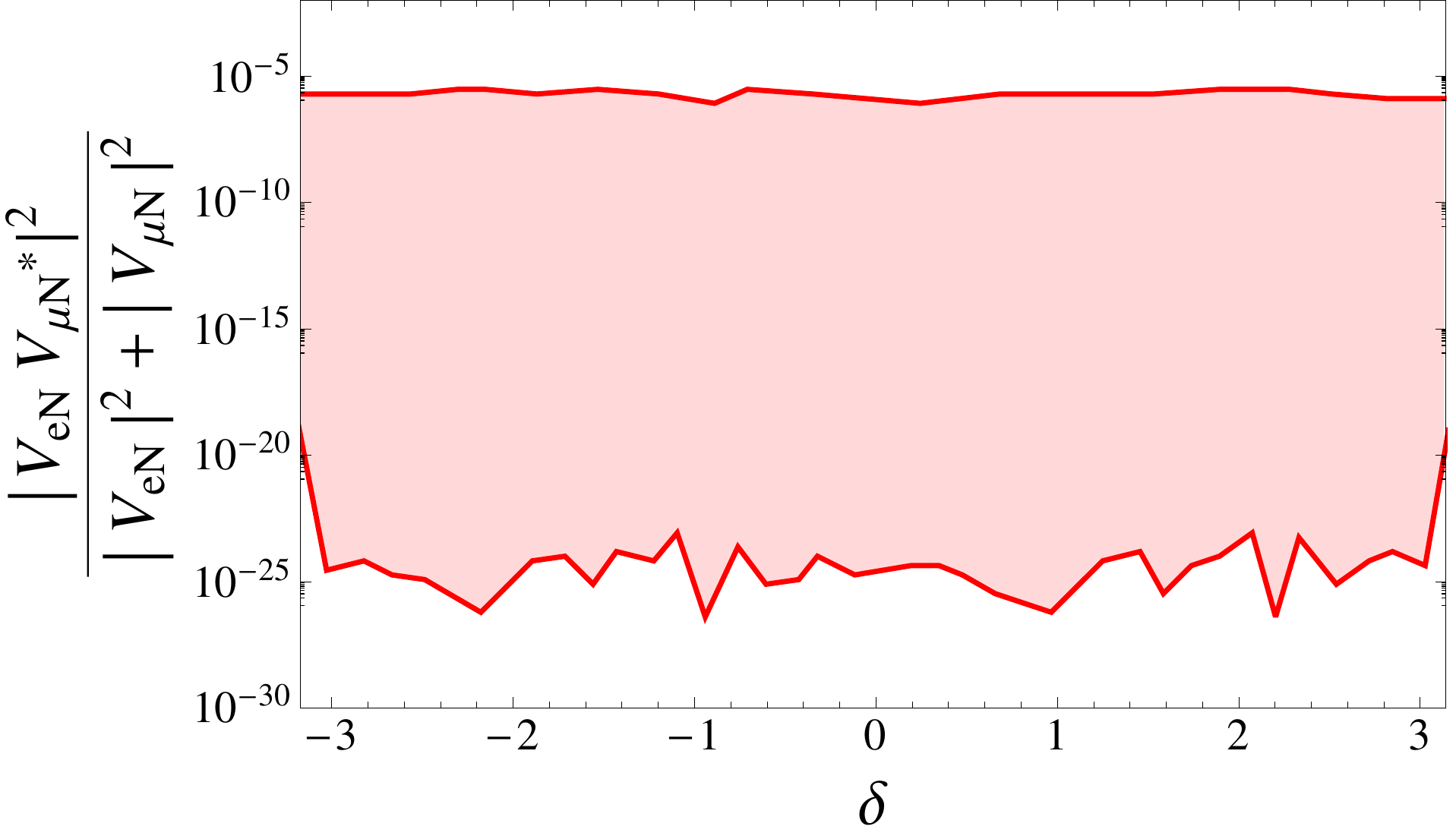}
\includegraphics[width=0.5\textwidth,angle=0,scale=0.9]{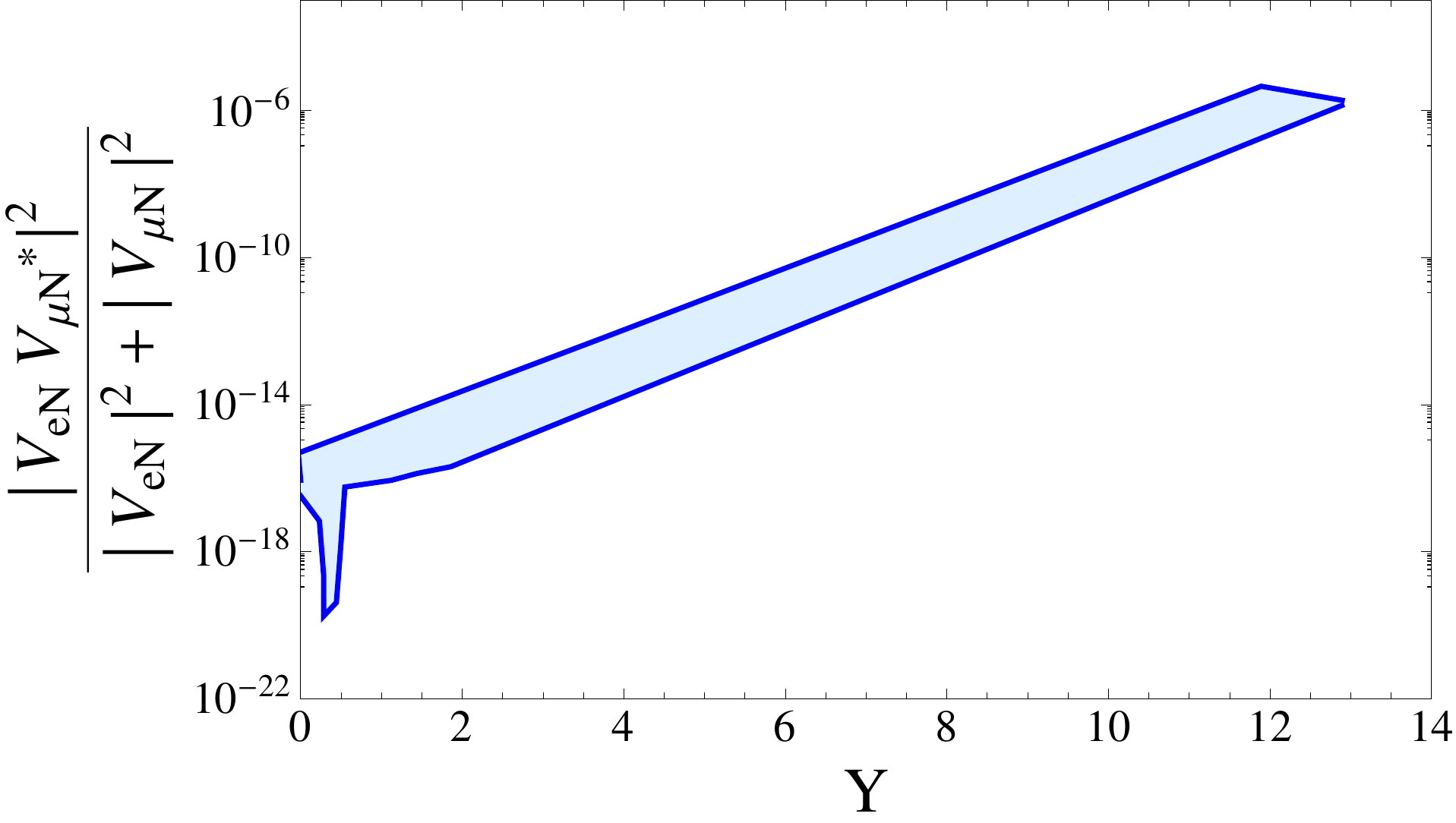}
\end{center}
\caption{
The allowed parameter region for a combination of the mixing parameters, 
  $ |V_{eN} V_{\mu N}^\ast|^2 /(|V_{eN}|^2+ |V_{\mu N}|^2)$, 
  in the NH case. 
}
\label{mixNHc}
\end{figure}
\begin{figure}
\begin{center}
\includegraphics[width=0.55\textwidth,angle=0,scale=0.9]{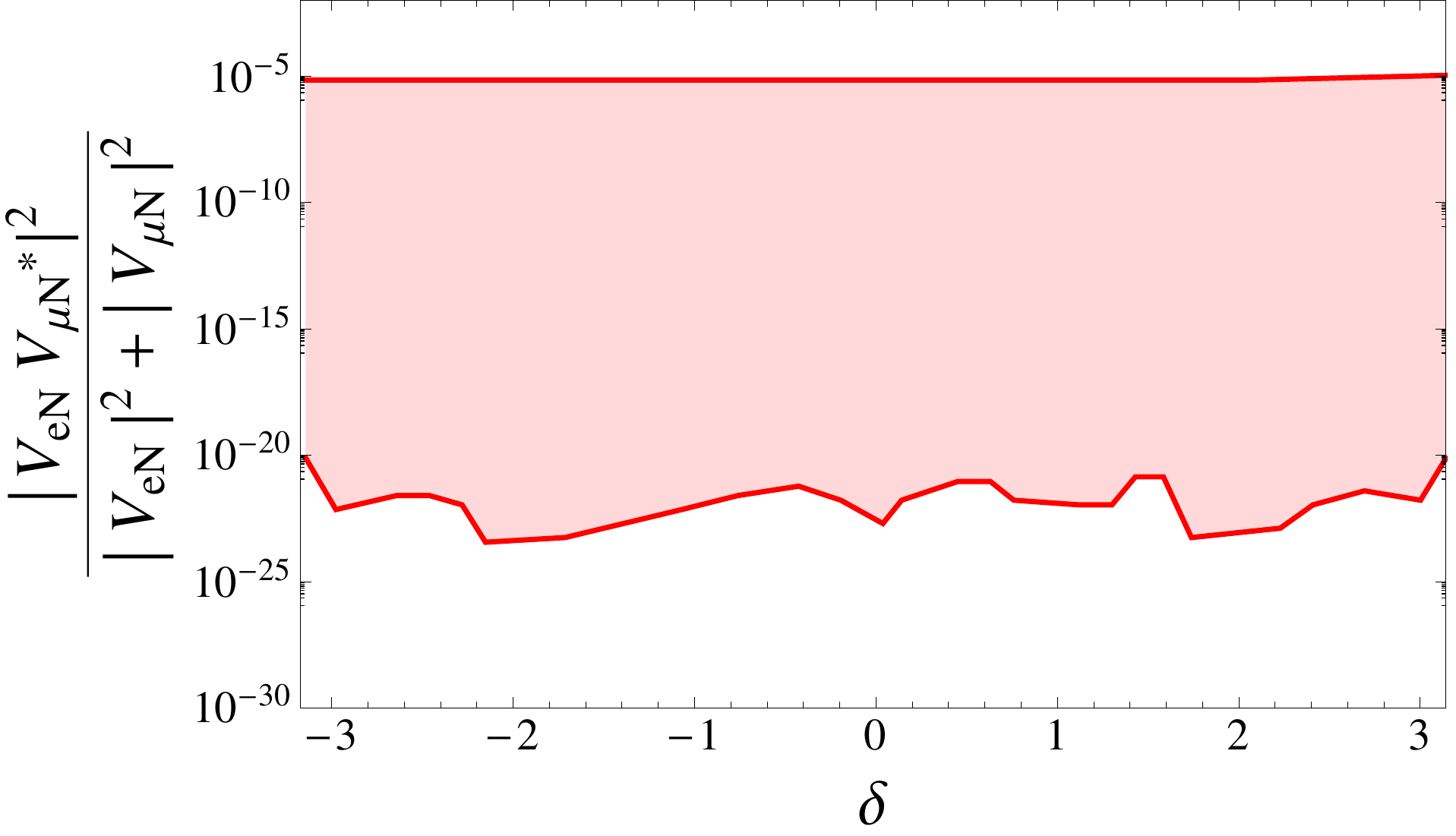}
\includegraphics[width=0.55\textwidth,angle=0,scale=0.9]{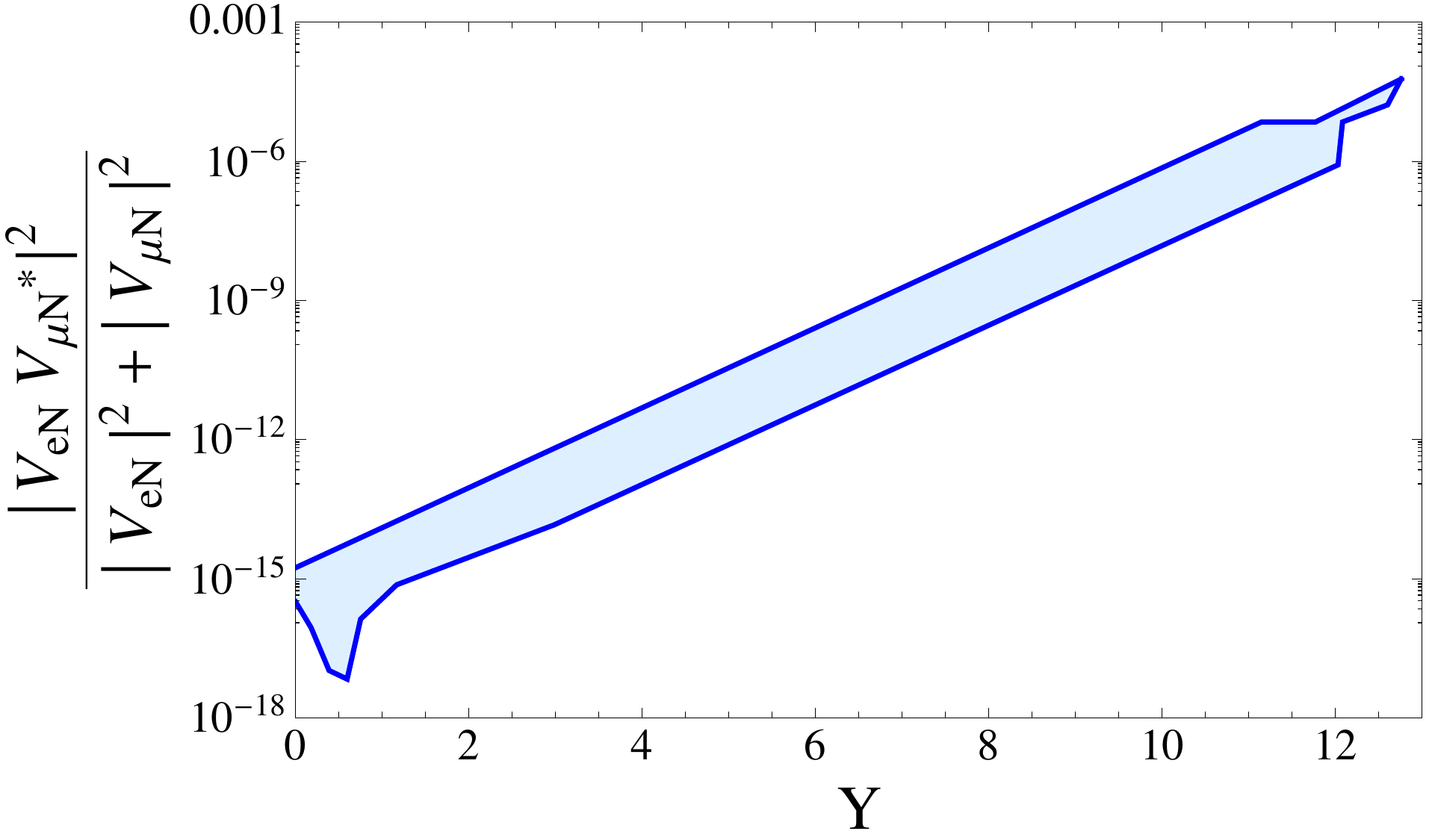}
\end{center}
\caption{
Same as Fig.~\ref{mixNHc} but for the IH case. 
}
\label{mixIHc}
\end{figure}

In our analysis, we set $M_N=100$ GeV and vary the three parameters 
   in the range of $-\pi \leq \delta, \rho \leq \pi$ with the interval of $\frac{\pi}{20}$ 
   and $0\leq y \leq 14$ with the interval of $0.01875$.
For the NH case, we show in Fig.~\ref{mixNH1a} our results on the mixing matrix element $|{\cal R}_{\alpha i}|^2$  
  with respect to $- \pi < \delta < \pi$. 
In each panel, the shaded region satisfies the experimental constraints in Eq.~(\ref{eps}). 
We have found $|{\cal R}_{\alpha i}|^2 < 2.94\times 10^{-4}$.  
Note that as in Eqs.~(\ref{X_LHC}) and (\ref{X_LC}), 
   the heavy neutrino production cross section 
   is proportional to $|{\cal R}_{\alpha i}|^2$ and hence the constraints in Eq.~(\ref{eps}) 
   provide us with the upper bound on the cross section. 
The same results but with respect to $Y$ are shown in  Fig.~\ref{mixNH1b}. 
For the IH case, the corresponding results are shown in Fig.~\ref{mixIH1a} and Fig.~\ref{mixIH1b}, respectively. 
Similarly to the NH case, we have found $|{\cal R}_{\alpha i}|^2 < 3.52\times10^{-4}$.  
We also show in Fig.~\ref{mixNHc} and Fig.~\ref{mixIHc} 
  our results for a combination of the mixing parameters, 
  $ |V_{eN} V_{\mu N}^\ast|^2/(|V_{eN}|^2+ |V_{\mu N}|^2)$, 
  in the NH and IH cases, respectively. 
For comparison, we list in Table~\ref{bounds} the upper bounds on the mixing parameters 
  from the collider experiments, for $M_N=100$ GeV. 
We can see that the upper bounds on the mixing we have obtained are more severe 
  than those listed in Table~\ref{bounds}.

\begin{table}[ht]
\begin{center}
\begin{tabular}{c|c|c }
\hline
\hline
      Experiments& Mixning angles&Upper Bounds \\
\hline
\hline
EWPD-e\cite{deBlas:2013gla, delAguila:2008pw, Akhmedov:2013hec}&$|V_{eN}|^2$&$1.7 \times 10^{-3}$\\
EWPD-$\mu$\cite{deBlas:2013gla, delAguila:2008pw, Akhmedov:2013hec}&$|V_{\mu N}|^2$&$9.0\times 10^{-3}$\\
EWPD-$\tau$\cite{deBlas:2013gla, delAguila:2008pw, Akhmedov:2013hec}&$|V_{\tau N}|^2$&$4.2 \times 10^{-3}$\\
\hline
\hline
L3\cite{Achard:2001qv}&$|V_{\ell N}|^2$, $\ell=e, \mu$&$2.2 \times 10^{-3}$\\
\hline
\hline
Higgs-LHC\cite{BhupalDev:2012zg}&$|V_{\ell N}|^2$, $\ell=e, \mu$&$3.4 \times10^{-3}$\\
\hline
\hline
LHC-e(ATLAS, 8 TeV)\cite{Aad:2015xaa} &$|V_{e N}|^{2}$&$4.1\times 10^{-2}$\\
LHC-$\mu$(ATLAS, 8 TeV)\cite{Aad:2015xaa} &$|V_{\mu N}|^{2}$&$1.9 \times 10^{-3}$\\
LHC-e(CMS, 8 TeV)\cite{CMS8:2016olu}&$|V_{eN}|^{2}$&$1.1 \times 10^{-2}$\\
LHC-$\mu$(CMS, 8 TeV)\cite{CMS8:2016olu}&$|V_{eN}|^{2}$&$4.6 \times 10^{-3}$\\
LHC-e, $\mu$(CMS, 8 TeV)\cite{CMS8:2016olu}&$\frac{|V_{eN} V_{\mu N}^\ast|^{2}}{|V_{eN}|^2+ |V_{\mu N}|^2}$&$2.4 \times 10^{-3}$\\
\hline
\hline
\end{tabular}
\end{center}
\caption{
Upper bounds on the mixing parameters for $M_{N}=100$ GeV in the type-I seesaw framework 
  from the various collider experiments. 
}
\label{bounds}
\end{table}

In summary, we have studied the minimal type-I seesaw scenario and the current experimental bounds 
  on the mixing between the heavy Majorana neutrinos and the SM neutrinos. 
We have employed the general parameterization for the neutrino Dirac mass matrix 
  so as to reproduce all neutrino oscillation data. 
In this way, the model is controlled by only three parameters, the Dirac $CP$-phase, one Majorana phase, 
  and the (complex) angle of the $2 \times 2$ orthogonal matrix 
  with the degenerate heavy neutrino mass $M_N=100$ GeV. 
We have performed the parameter scan to identify the allowed parameter region 
  which satisfies the experimental constraints from 
  the electroweak precision measurements and the lepton-flavor violations.     
For the allowed parameter region, we have found the upper bound on the mixing parameters 
  to be $|{\cal R}_{\alpha i}|^2 \lesssim 10^{-4}$, which is more severe than those obtained 
  from the search for heavy Majorana neutrinos at the current LHC experiments. 
The region $|{\cal R}_{\alpha i}|^2 \lesssim 10^{-4}$ we have found 
  can be tested at the High-Luminosity LHC or at a 100 TeV pp-collider in the future. 
We have also performed parameter scan for the effective neutrino mass relevant to the neutrinoless double beta decay 
and found the range of $0.00154 \leq |m^\nu_{ee}| ({\rm eV})\leq 0.00389$ (NH case) and $0.0167 \leq |m^\nu_{ee}| ({\rm eV})\leq 0.0473$
(IH case), which are consistent with the current experimental bound $\lesssim 0.1$ eV \cite{KamLAND-Zen:2016pfg}.

From Figs.~\ref{mixNH1b} and \ref{mixIH1b}, we can see that the upper bounds on the mixing parameters 
  are obtained for $Y \sim 12$. 
For such a $Y$ value, the matrix in Eq.~(\ref{OO}) is approximately proportional to $e^{2 Y}$, and hence 
  $\epsilon \propto e^{2 Y}/M_N$ in Eq.~(\ref{RR}) and  
  the upper bound on $e^{2 Y}/M_N$ is determined from the constraint of Eq~(\ref{eps}). 
In this case, the mixing matrix is roughly proportional to $e^{Y}/\sqrt{M_N}=\sqrt{e^{2 Y}/M_N}$ 
  and its upper bound is fixed accordingly. 
Although the value of $Y$ to yield the upper bound is a function of $M_N$,  
  the upper bounds on the mixing matrix elements are almost independent of $M_N$.  
However, the cross section of the heavy neutrino at the LHC is exponentially decreasing 
  as $M_N$ values are increased, because of the energy dependence of the parton distribution functions.

Although we have shown the results only for the case with the degenerate heavy neutrinos, 
  we have also performed parameter scans for the non-generate case 
  with a few sample values of $m_N^{~2} > m_N^{~1}=100$ GeV 
  and found that the upper bound on the mixing parameters reduces 
  from the case with $m_N^{~2} = m_N^{~1}=100$ GeV. 
This observation suggests that the degenerate mass spectrum is preferable 
  in terms of the testability of the type-I seesaw scenario at the future collider experiments. 
Our parameter scan analysis in this letter is similar to that in Ref.~\cite{Das:2012ze}, 
  where the inverse-seesaw scenario was considered. 
A crucial difference of the inverse-seesaw scenario is that we can choose 
  a flavor-blind Dirac mass matrix by encoding all the flavor structures into the small lepton-number 
  violating parameter $\mu_{ij}$ and easily avoid the experimental constraints in Eq.~(\ref{eps}). 
However, there is no such freedom in the type-I seesaw scenario, and the neutrino Dirac mass matrix 
  must satisfy all the experimental data such as the neutrino oscillation data, 
  the electroweak precision measurements, and the lepton-flavor violations. 
As a result, the heavy neutrino production cross section at the high energy colliders 
  are constrained very severely.

\acknowledgments
The authors would like to thank P.~S.~Bhupal Dev for useful comments. 
A.D. would like to thank Bose Institute, Kolkata for hospitality and arranging an academic visit where a part of the work was done.
 The work of A.D. is supported by the Korea Neutrino Research Center which 
  is established by the National Research Foundation of Korea (NRF) grant funded by the Korea government (MSIP) (No.~2009-0083526). 
The work of N.O. is supported in part by the United States Department of Energy (No.~DE-SC0013680).



\begin{thebibliography}{999}
\bibitem{Neut1}
K.~Abe et.~al. [T2K Collaboration] Phys.~Rev.~Lett.~107, 041801 (2011).
\bibitem{Neut2}
P.~Adamson et al. [MINOS Collaboration], Phys.~Rev.~Lett.~107, 181802 (2011).
\bibitem{Neut4}
Y.~Abe et al. [DOUBLE-CHOOZ Collaboration], Phys.~Rev.~Lett.~108, 131801 (2012).
\bibitem{Neut5}
F.~P.~An et al. [DAYA-BAY Collaboration], Phys.~Rev.~Lett.~108, 171803 (2012).
\bibitem{Neut6}
J.~K.~Ahn et al. [RENO Collaboration], Phys.~Rev.~Lett.~108, 191802 (2012).    
\bibitem{Neut3}
C.~Patrignani {\it et al.} [Particle Data Group],
  ``Review of Particle Physics,''
  Chin.\ Phys.\ C {\bf 40}, no. 10, 100001 (2016).
  doi:10.1088/1674-1137/40/10/100001

  \bibitem{seesaw0} 
  P.~Minkowski,
  ``$\mu \to e\gamma$ at a Rate of One Out of $10^{9}$ Muon Decays?,''
  Phys.\ Lett.\  {\bf 67B}, 421 (1977).
  doi:10.1016/0370-2693(77)90435-X  
  \bibitem{seesaw1}
T.~Yanagida, ``Horizontal Symmetry and Masses of Neutrinos,'' Prog.~Theor.~Phys.~{\bf 64}, 1103 (1980).
\bibitem{seesaw2}
J.~Schechter and J.~W.~F.~Valle, ``Neutrino Masses in SU(2) $\otimes$ U(1) Theories,'' Phys.~Rev.~D {\bf 22}, 2227 (1980). 
\bibitem{seesaw3}
T.~Yanagida, in Proceedings of the Work- shop on the Unified Theory and the Baryon Number in the Universe (O.~Sawada and A.~Sugamoto, eds.), KEK, Tsukuba, Japan, 1979, p.~95.
\bibitem{seesaw4}
M.~Gell-Mann, P. Ramond, and R.~Slansky, Supergravity (P.~van Nieuwenhuizen et al.~eds.), North Holland, Amsterdam, 1979, p.~315.
\bibitem{seesaw5}
S.~L.~Glashow, The future of elementary particle physics, in Proceedings of the 1979 Carg`ese Summer Institute on Quarks and Leptons (M. Levy et al. eds.), Plenum Press, New York, 1980, p.~687.
\bibitem{seesaw6}
R.~N.~Mohapatra and G.~Senjanovic, ``Neutrino Mass and Spontaneous Parity Violation,'' Phys.~Rev.~Lett.~{\bf 44}, 912 (1980).    
\bibitem{Weinberg:1979sa} 
  S.~Weinberg,
  ``Baryon and Lepton Nonconserving Processes,''
  Phys.\ Rev.\ Lett.\  {\bf 43}, 1566 (1979).
  doi:10.1103/PhysRevLett.43.1566  
\bibitem{Escrihuela:2015wra} 
  F.~J.~Escrihuela, D.~V.~Forero, O.~G.~Miranda, M.~Tortola and J.~W.~F.~Valle,
  ``On the description of nonunitary neutrino mixing,''
  Phys.\ Rev.\ D {\bf 92}, no. 5, 053009 (2015)
  Erratum: [Phys.\ Rev.\ D {\bf 93}, no. 11, 119905 (2016)]
  doi:10.1103/PhysRevD.93.119905, 10.1103/PhysRevD.92.053009
  [arXiv:1503.08879 [hep-ph]].


\bibitem{Ge:2016xya} 
  S.~F.~Ge, P.~Pasquini, M.~Tortola and J.~W.~F.~Valle,
  ``Measuring the leptonic CP phase in neutrino oscillations with nonunitary mixing,''
  Phys.\ Rev.\ D {\bf 95}, no. 3, 033005 (2017)
  doi:10.1103/PhysRevD.95.033005
  [arXiv:1605.01670 [hep-ph]].  

  \bibitem{Constraints1}
  S.~Antusch, C.~Biggio, E.~Fernandez-Martinez, M.~B.~Gavela and J.~Lopez-Pavon,
  ``Unitarity of the Leptonic Mixing Matrix,''
  JHEP {\bf 0610}, 084 (2006).
  [arXiv:hep-ph/0607020].


\bibitem{Constraints2}
  A.~Abada, C.~Biggio, F.~Bonnet, M.~B.~Gavela and T.~Hambye,
  ``Low energy effects of neutrino masses,''
  JHEP {\bf 0712}, 061 (2007).
    [arXiv:0707.4058 [hep-ph]].


\bibitem{Constraints3}
A.~Ibarra, E.~Molinaro and S.~T.~Petcov,
 ``TeV Scale See-Saw Mechanisms of Neutrino Mass Generation, the Majorana Nature of the Heavy Singlet Neutrinos and $(\beta\beta)_{0\nu}$-Decay,''
  JHEP {\bf 1009}, 108 (2010); 
  [arXiv:1007.2378 [hep-ph]].  
  
  
  \bibitem{Constraints4} 
    A.~Ibarra, E.~Molinaro and S.~T.~Petcov,
``Low Energy Signatures of the TeV Scale See-Saw Mechanism,''
  Phys.\ Rev.\ D {\bf 84}, 013005 (2011); 
 [arXiv:1103.6217 [hep-ph]].


\bibitem{Constraints5}
D.~N.~Dinh, A.~Ibarra, E.~Molinaro and S.~T.~Petcov,
  ``The $\mu - e$ Conversion in Nuclei, $\mu \to e \gamma, \mu \to 3e$ Decays and TeV Scale See-Saw Scenarios of Neutrino Mass Generation,''
  JHEP {\bf 1208}, 125 (2012)
  [Erratum-ibid.\  {\bf 1309}, 023 (2013)].
   [arXiv:1205.4671 [hep-ph]].


\bibitem{Asaka:2011pb} 
  T.~Asaka, S.~Eijima and H.~Ishida,
  ``Mixing of Active and Sterile Neutrinos,''
  JHEP {\bf 1104}, 011 (2011)
  doi:10.1007/JHEP04(2011)011
  [arXiv:1101.1382 [hep-ph]].
  

\bibitem{Ruchayskiy:2011aa} 
  O.~Ruchayskiy and A.~Ivashko,
  ``Experimental bounds on sterile neutrino mixing angles,''
  JHEP {\bf 1206}, 100 (2012)
  doi:10.1007/JHEP06(2012)100
  [arXiv:1112.3319 [hep-ph]].

  
\bibitem{Gorbunov:2014ypa} 
  D.~Gorbunov and I.~Timiryasov,
  ``Testing $\nu$MSM with indirect searches,''
  Phys.\ Lett.\ B {\bf 745}, 29 (2015)
  doi:10.1016/j.physletb.2015.02.060
  [arXiv:1412.7751 [hep-ph]].    
  
\bibitem{Drewes:2015iva} 
  M.~Drewes and B.~Garbrecht,
  ``Combining experimental and cosmological constraints on heavy neutrinos,''
  Nucl.\ Phys.\ B {\bf 921}, 250 (2017)
  doi:10.1016/j.nuclphysb.2017.05.001
  [arXiv:1502.00477 [hep-ph]].  
  

\bibitem{Hernandez:2016kel} 
  P.~Hernandez, M.~Kekic, J.~Lopez-Pavon, J.~Racker and J.~Salvado,
  ``Testable Baryogenesis in Seesaw Models,''
  JHEP {\bf 1608}, 157 (2016)
  doi:10.1007/JHEP08(2016)157
  [arXiv:1606.06719 [hep-ph]].

  
 \bibitem{Drewes:2016jae} 
  M.~Drewes, B.~Garbrecht, D.~Gueter and J.~Klaric,
  ``Testing the low scale seesaw and leptogenesis,''
  arXiv:1609.09069 [hep-ph].


\bibitem{Abada:2013aba} 
  A.~Abada, A.~M.~Teixeira, A.~Vicente and C.~Weiland,
  ``Sterile neutrinos in leptonic and semileptonic decays,''
  JHEP {\bf 1402}, 091 (2014)
  doi:10.1007/JHEP02(2014)091
  [arXiv:1311.2830 [hep-ph]].
  
\bibitem{Abada:2014kba} 
  A.~Abada, M.~E.~Krauss, W.~Porod, F.~Staub, A.~Vicente and C.~Weiland,
  ``Lepton flavor violation in low-scale seesaw models: SUSY and non-SUSY contributions,''
  JHEP {\bf 1411}, 048 (2014)
  doi:10.1007/JHEP11(2014)048
  [arXiv:1408.0138 [hep-ph]].   
  
\bibitem{Asaka:2014kia} 
  T.~Asaka, S.~Eijima and K.~Takeda,
  ``Lepton Universality in the $\nu$MSM,''
  Phys.\ Lett.\ B {\bf 742}, 303 (2015)
  doi:10.1016/j.physletb.2015.01.049
  [arXiv:1410.0432 [hep-ph]].  

\bibitem{Fernandez-Martinez:2015hxa} 
  E.~Fernandez-Martinez, J.~Hernandez-Garcia, J.~Lopez-Pavon and M.~Lucente,
  ``Loop level constraints on Seesaw neutrino mixing,''
  JHEP {\bf 1510}, 130 (2015)
  doi:10.1007/JHEP10(2015)130
  [arXiv:1508.03051 [hep-ph]]. 

\bibitem{deGouvea:2015euy} 
  A.~de Gouvea and A.~Kobach,
  ``Global Constraints on a Heavy Neutrino,''
  Phys.\ Rev.\ D {\bf 93}, no. 3, 033005 (2016)
  doi:10.1103/PhysRevD.93.033005
  [arXiv:1511.00683 [hep-ph]].
  
  
\bibitem{Fernandez-Martinez:2016lgt} 
  E.~Fernandez-Martinez, J.~Hernandez-Garcia and J.~Lopez-Pavon,
  ``Global constraints on heavy neutrino mixing,''
  JHEP {\bf 1608}, 033 (2016)
  doi:10.1007/JHEP08(2016)033
  [arXiv:1605.08774 [hep-ph]].



\bibitem{minimal_seesaw1}
S.~F.~King,
  ``Large mixing angle MSW and atmospheric neutrinos from single right-handed neutrino dominance and U(1) family symmetry,''
  Nucl.\ Phys.\ B {\bf 576}, 85 (2000)
  doi:10.1016/S0550-3213(00)00109-7
  [hep-ph/9912492].
\bibitem{minimal_seesaw2}
P.~H.~Frampton, S.~L.~Glashow and T.~Yanagida,
  ``Cosmological sign of neutrino CP violation,''
  Phys.\ Lett.\ B {\bf 548}, 119 (2002)
  doi:10.1016/S0370-2693(02)02853-8
  [hep-ph/0208157].
 \bibitem{Das:2012ze} 
  A.~Das and N.~Okada,
  ``Inverse seesaw neutrino signatures at the LHC and ILC,''
  Phys.\ Rev.\ D {\bf 88}, 113001 (2013)
  doi:10.1103/PhysRevD.88.113001
  [arXiv:1207.3734 [hep-ph]].  
   \bibitem{Dev:2013wba} 
  P.~S.~B.~Dev, A.~Pilaftsis and U.~k.~Yang,
  ``New Production Mechanism for Heavy Neutrinos at the LHC,''
  Phys.\ Rev.\ Lett.\  {\bf 112}, no. 8, 081801 (2014)
  doi:10.1103/PhysRevLett.112.081801
  [arXiv:1308.2209 [hep-ph]].  
    \bibitem{Das:2014jxa} 
  A.~Das, P.~S.~Bhupal Dev and N.~Okada,
  ``Direct bounds on electroweak scale pseudo-Dirac neutrinos from $\sqrt s=8$ TeV LHC data,''
  Phys.\ Lett.\ B {\bf 735}, 364 (2014)
  doi:10.1016/j.physletb.2014.06.058
  [arXiv:1405.0177 [hep-ph]].    
    \bibitem{Alva:2014gxa} 
  D.~Alva, T.~Han and R.~Ruiz,
  ``Heavy Majorana neutrinos from $W\gamma$ fusion at hadron colliders,''
  JHEP {\bf 1502}, 072 (2015)
  doi:10.1007/JHEP02(2015)072
  [arXiv:1411.7305 [hep-ph]].      
   \bibitem{Das:2015toa} 
  A.~Das and N.~Okada,
  ``Improved bounds on the heavy neutrino productions at the LHC,''
  Phys.\ Rev.\ D {\bf 93}, no. 3, 033003 (2016)
  doi:10.1103/PhysRevD.93.033003
  [arXiv:1510.04790 [hep-ph]].
  \bibitem{Hessler:2014ssa} 
  A.~G.~Hessler, A.~Ibarra, E.~Molinaro and S.~Vogl,
  ``Impact of the Higgs boson on the production of exotic particles at the LHC,''
  Phys.\ Rev.\ D {\bf 91}, no. 11, 115004 (2015)
  doi:10.1103/PhysRevD.91.115004
  [arXiv:1408.0983 [hep-ph]].  
  \bibitem{Degrande:2016aje} 
  C.~Degrande, O.~Mattelaer, R.~Ruiz and J.~Turner,
  ``Fully-Automated Precision Predictions for Heavy Neutrino Production Mechanisms at Hadron Colliders,''
  Phys.\ Rev.\ D {\bf 94}, no. 5, 053002 (2016)
  doi:10.1103/PhysRevD.94.053002
  [arXiv:1602.06957 [hep-ph]].    
   \bibitem{Das:2016hof} 
  A.~Das, P.~Konar and S.~Majhi,
  ``Production of Heavy neutrino in next-to-leading order QCD at the LHC and beyond,''
  JHEP {\bf 1606}, 019 (2016)
  doi:10.1007/JHEP06(2016)019
  [arXiv:1604.00608 [hep-ph]].  
\bibitem{Das:2017pvt} 
  A.~Das,
  ``Pair production of heavy neutrinos in next-to-leading order QCD at the hadron colliders in the inverse seesaw framework,''
  arXiv:1701.04946 [hep-ph].
   \bibitem{Antusch:2014woa}
  S.~Antusch and O.~Fischer,
  ``Non-unitarity of the leptonic mixing matrix: Present bounds and future sensitivities,''
  JHEP {\bf 1410} (2014) 094
  doi:10.1007/JHEP10(2014)094
  [arXiv:1407.6607 [hep-ph]].  
   \bibitem{Antusch:2015mia}
  S.~Antusch and O.~Fischer,
  ``Testing sterile neutrino extensions of the Standard Model at future lepton colliders,''
  JHEP {\bf 1505} (2015) 053
  doi:10.1007/JHEP05(2015)053
  [arXiv:1502.05915 [hep-ph]].  
  \bibitem{Antusch:2015rma} 
  S.~Antusch and O.~Fischer,
  ``Testing sterile neutrino extensions of the Standard Model at the Circular Electron Positron Collider,''
  Int.\ J.\ Mod.\ Phys.\ A {\bf 30}, no. 23, 1544004 (2015).
  doi:10.1142/S0217751X15440042  
  \bibitem{Antusch:2015gjw} 
  S.~Antusch, E.~Cazzato and O.~Fischer,
  ``Higgs production from sterile neutrinos at future lepton colliders,''
  JHEP {\bf 1604}, 189 (2016)
  doi:10.1007/JHEP04(2016)189
  [arXiv:1512.06035 [hep-ph]].  
   \bibitem{Antusch:2016brq} 
  S.~Antusch and O.~Fischer,
  ``Probing the nonunitarity of the leptonic mixing matrix at the CEPC,''
  Int.\ J.\ Mod.\ Phys.\ A {\bf 31}, no. 33, 1644006 (2016)
  doi:10.1142/S0217751X16440061
  [arXiv:1604.00208 [hep-ph]].  
  \bibitem{Antusch:2016vyf} 
  S.~Antusch, E.~Cazzato and O.~Fischer,
  ``Displaced vertex searches for sterile neutrinos at future lepton colliders,''
  JHEP {\bf 1612}, 007 (2016)
  doi:10.1007/JHEP12(2016)007
  [arXiv:1604.02420 [hep-ph]].  
  \bibitem{Fischer:2016rsh} 
  O.~Fischer,
  ``Clues on the Majorana scale from scalar resonances at the LHC,''
  arXiv:1607.00282 [hep-ph].  
  \bibitem{Antusch:2016qby} 
  S.~Antusch, E.~Cazzato and O.~Fischer,
  ``Higgs production through sterile neutrinos,''
  Int.\ J.\ Mod.\ Phys.\ A {\bf 31}, no. 33, 1644007 (2016).
  doi:10.1142/S0217751X16440073
  \bibitem{Antusch:2016ejd} 
  S.~Antusch, E.~Cazzato and O.~Fischer,
  ``Sterile neutrino searches at future $e^-e^+$, $pp$, and $e^-p$ colliders,''
  arXiv:1612.02728 [hep-ph].  
     \bibitem{Dib:2015oka} 
  C.~O.~Dib and C.~S.~Kim,
  ``Discovering sterile Neutrinos ligther than $M_W$ at the LHC,''
  Phys.\ Rev.\ D {\bf 92}, no. 9, 093009 (2015)
  doi:10.1103/PhysRevD.92.093009
  [arXiv:1509.05981 [hep-ph]].     
  
     
     \bibitem{Dib:2016wge} 
  C.~O.~Dib, C.~S.~Kim, K.~Wang and J.~Zhang,
  ``Distinguishing Dirac/Majorana Sterile Neutrinos at the LHC,''
  Phys.\ Rev.\ D {\bf 94}, no. 1, 013005 (2016)
  doi:10.1103/PhysRevD.94.013005
  [arXiv:1605.01123 [hep-ph]]     
  

\bibitem{Asaka:2013jfa} 
  T.~Asaka and S.~Eijima,
  ``Direct Search for Right-handed Neutrinos and Neutrinoless Double Beta Decay,''
  PTEP {\bf 2013}, no. 11, 113B02 (2013)
  doi:10.1093/ptep/ptt094
  [arXiv:1308.3550 [hep-ph]].
  


  \bibitem{Dib:2014iga} 
  C.~Dib and C.~S.~Kim,
  ``Remarks on the lifetime of sterile neutrinos and the effect on detection of rare meson decays $M^+ \to M^{\prime}-\ell^+\ell^+$,''
  Phys.\ Rev.\ D {\bf 89}, no. 7, 077301 (2014)
  doi:10.1103/PhysRevD.89.077301
  [arXiv:1403.1985 [hep-ph]].  
  
  
  \bibitem{Dib:2014pga} 
  C.~O.~Dib, M.~Campos and C.~S.~Kim,
  ``CP Violation with Majorana neutrinos in K Meson Decays,''
  JHEP {\bf 1502}, 108 (2015)
  doi:10.1007/JHEP02(2015)108
  [arXiv:1403.8009 [hep-ph]].  
  
  \bibitem{Cvetic:2012hd} 
  G.~Cvetic, C.~Dib and C.~S.~Kim,
  ``Probing Majorana neutrinos in rare $\pi^+ \to e^+ e^+ \mu^- \nu$ decays,''
  JHEP {\bf 1206}, 149 (2012)
  doi:10.1007/JHEP06(2012)149
  [arXiv:1203.0573 [hep-ph]].  
  
  \bibitem{Cvetic:2010rw} 
  G.~Cvetic, C.~Dib, S.~K.~Kang and C.~S.~Kim,
 ``Probing Majorana neutrinos in rare K and D,$ \sim D_s$, B, $B_c$ meson decays,''
  Phys.\ Rev.\ D {\bf 82}, 053010 (2010)
  doi:10.1103/PhysRevD.82.053010
  [arXiv:1005.4282 [hep-ph]].  
   
   
 \bibitem{Cvetic:2016fbv} 
  G.~Cvetic and C.~S.~Kim,
  ``Rare decays of B mesons via on-shell sterile neutrinos,''
  Phys.\ Rev.\ D {\bf 94}, no. 5, 053001 (2016)
  doi:10.1103/PhysRevD.94.053001
  [arXiv:1606.04140 [hep-ph]].   
  

\bibitem{Zamora-Saa:2016ito} 
  J.~Zamora-Saa,
  ``Resonant $CP$ violation in rare $\tau^{\pm}$ decays,''
  JHEP {\bf 1705}, 110 (2017)
  doi:10.1007/JHEP05(2017)110
  [arXiv:1612.07656 [hep-ph]].
 
  
  \bibitem{Rasmussen:2016njh} 
  R.~W.~Rasmussen and W.~Winter,
  ``Perspectives for tests of neutrino mass generation at the GeV scale: Experimental reach versus theoretical predictions,''
  Phys.\ Rev.\ D {\bf 94}, no. 7, 073004 (2016)
  doi:10.1103/PhysRevD.94.073004
  [arXiv:1607.07880 [hep-ph]].
  \bibitem{Bambhaniya:2014hla} 
  G.~Bambhaniya, S.~Khan, P.~Konar and T.~Mondal,
  ``Constraints on a seesaw model leading to quasidegenerate neutrinos and signatures at the LHC,''
  Phys.\ Rev.\ D {\bf 91}, no. 9, 095007 (2015)
  doi:10.1103/PhysRevD.91.095007
  [arXiv:1411.6866 [hep-ph]].  
  \bibitem{Bambhaniya:2014kga} 
  G.~Bambhaniya, S.~Goswami, S.~Khan, P.~Konar and T.~Mondal,
  ``Looking for hints of a reconstructible seesaw model at the Large Hadron Collider,''
  Phys.\ Rev.\ D {\bf 91}, 075007 (2015)
  doi:10.1103/PhysRevD.91.075007
  [arXiv:1410.5687 [hep-ph]].  
  \bibitem{Bambhaniya:2016rbb} 
  G.~Bambhaniya, P.~S.~B.~Dev, S.~Goswami, S.~Khan and W.~Rodejohann,
  ``Naturalness, Vacuum Stability and Leptogenesis in the Minimal Seesaw Model,''
  arXiv:1611.03827 [hep-ph].  
 \bibitem{Blennow:2016jkn} 
  M.~Blennow, P.~Coloma, E.~Fernandez-Martinez, J.~Hernandez-Garcia and J.~Lopez-Pavon,
  ``Non-Unitarity, sterile neutrinos, and Non-Standard neutrino Interactions,''
  arXiv:1609.08637 [hep-ph].
  \bibitem{Caputo:2016ojx} 
  A.~Caputo, P.~Hernandez, M.~Kekic, J.~Lopez-Pavon and J.~Salvado,
  ``The seesaw path to leptonic CP violation,''
  arXiv:1611.05000 [hep-ph].  
  \bibitem{Deppisch:2015qwa} 
  F.~F.~Deppisch, P.~S.~Bhupal Dev and A.~Pilaftsis,
  ``Neutrinos and Collider Physics,''
  New J.\ Phys.\  {\bf 17}, no. 7, 075019 (2015)
  doi:10.1088/1367-2630/17/7/075019
  [arXiv:1502.06541 [hep-ph]].  
  \bibitem{Dev:2014xea} 
  P.~S.~Bhupal Dev, S.~Goswami and M.~Mitra,
  ``TeV Scale Left-Right Symmetry and Large Mixing Effects in Neutrinoless Double Beta Decay,''
  Phys.\ Rev.\ D {\bf 91}, no. 11, 113004 (2015)
  doi:10.1103/PhysRevD.91.113004
  [arXiv:1405.1399 [hep-ph]].
\bibitem{CI_Para}
J.~A.~Casas and A.~Ibarra,
  ``Oscillating neutrinos and muon $\to$ e, gamma,''
  Nucl.\ Phys.\ B {\bf 618}, 171 (2001)
  doi:10.1016/S0550-3213(01)00475-8
  [hep-ph/0103065].


\bibitem{Lopez-Pavon:2015cga} 
  J.~Lopez-Pavon, E.~Molinaro and S.~T.~Petcov,
  ``Radiative Corrections to Light Neutrino Masses in Low Scale Type I Seesaw Scenarios and Neutrinoless Double Beta Decay,''
  JHEP {\bf 1511}, 030 (2015)
  doi:10.1007/JHEP11(2015)030
  [arXiv:1506.05296 [hep-ph]].


  
\bibitem{Adam}
J.~Adam et. al. [MEG Collaboration], 
Phys.\ Rev.\ Lett. {\bf 107},171801, (2011). 
 [arXiv: 1107.5541 [hep-ex]]

\bibitem{Aubert}
B.~Aubert et. al. [BABAR Collaboration],  
Phys.\ Rev.\ Lett. {\bf 104},021802,(2010). 
[arXiv: 0908.2381[hep-ex]]

\bibitem{OLeary}
See, for summary, 
 B.~O' Leary et. al. [SuperB Collaboration], 
 arXiv: 1008.1541[hep-ex].   
 \bibitem{deBlas:2013gla} 
  J.~de Blas,
  ``Electroweak limits on physics beyond the Standard Model,''
  EPJ Web Conf.\  {\bf 60}, 19008 (2013)
  doi:10.1051/epjconf/20136019008
  [arXiv:1307.6173 [hep-ph]]. 
  
 \bibitem{delAguila:2008pw} 
  F.~del Aguila, J.~de Blas and M.~Perez-Victoria,
  ``Effects of new leptons in Electroweak Precision Data,''
  Phys.\ Rev.\ D {\bf 78}, 013010 (2008)
  doi:10.1103/PhysRevD.78.013010
  [arXiv:0803.4008 [hep-ph]].  

 \bibitem{Akhmedov:2013hec} 
  E.~Akhmedov, A.~Kartavtsev, M.~Lindner, L.~Michaels and J.~Smirnov,
  ``Improving Electro-Weak Fits with TeV-scale Sterile Neutrinos,''
  JHEP {\bf 1305}, 081 (2013)
  doi:10.1007/JHEP05(2013)081
  [arXiv:1302.1872 [hep-ph]].  

 \bibitem{Achard:2001qv} 
  P.~Achard {\it et al.} [L3 Collaboration],
  ``Search for heavy isosinglet neutrino in $e^{+} e^{-}$ annihilation at LEP,''
  Phys.\ Lett.\ B {\bf 517}, 67 (2001)
  doi:10.1016/S0370-2693(01)00993-5
  [hep-ex/0107014].  

\bibitem{BhupalDev:2012zg} 
  P.~S.~Bhupal Dev, R.~Franceschini and R.~N.~Mohapatra,
  ``Bounds on TeV Seesaw Models from LHC Higgs Data,''
  Phys.\ Rev.\ D {\bf 86}, 093010 (2012)
  doi:10.1103/PhysRevD.86.093010
  [arXiv:1207.2756 [hep-ph]].
\bibitem{Aad:2015xaa} 
  G.~Aad {\it et al.} [ATLAS Collaboration],
  ``Search for heavy Majorana neutrinos with the ATLAS detector in pp collisions at $ \sqrt{s}=8 $ TeV,''
  JHEP {\bf 1507}, 162 (2015)
  doi:10.1007/JHEP07(2015)162
  [arXiv:1506.06020 [hep-ex]].
\bibitem{CMS8:2016olu}
V.~Khachatryan {\it et. al.} [CMS Collaboration],
``Search for heavy Majorana neutrinos in e$^{\pm}$e$^{\pm}$+jets and e$^{\pm}\mu^{\pm}$+jets events in proton-proton collisions at $\sqrt{s}=8$ TeV,"
JHEP {\bf 1604}, 169 (2016)
  doi: 10.1007/JHEP04 (2016)169
  [arxiv:1603.02248 [hep-ex]].
  
  \bibitem{KamLAND-Zen:2016pfg} 
  A.~Gando {\it et al.} [KamLAND-Zen Collaboration],
  ``Search for Majorana Neutrinos near the Inverted Mass Hierarchy Region with KamLAND-Zen,''
  Phys.\ Rev.\ Lett.\  {\bf 117}, no. 8, 082503 (2016)
  Addendum: [Phys.\ Rev.\ Lett.\  {\bf 117}, no. 10, 109903 (2016)]
  doi:10.1103/PhysRevLett.117.109903, 10.1103/PhysRevLett.117.082503
  [arXiv:1605.02889 [hep-ex]] .
   \end{thebibliography}
   \end{document}